\documentclass[aps,prb,superscriptaddress,showpacs,floatfix]{revtex4-1}
\usepackage{amsfonts}
\usepackage{graphicx,amsmath,amssymb,xspace,epsfig,float,multirow,subfigure,tabularx}

\begin{document}

\title{Mean field theory of short range order in strongly correlated low
dimensional electronic systems}
\author{Baruch Rosenstein }
\email{baruchro@hotmail.com}
\affiliation{Electrophysics Department, National Chiao Tung University, Hsinchu 30050,
\textit{Taiwan, R. O. C}}
\author{Dingping Li }
\email{lidp@pku.edu.cn}
\affiliation{School of Physics, Peking University, Beijing 100871, \textit{China}}
\affiliation{Collaborative Innovation Center of Quantum Matter, Beijing, China}
\author{Tianxing Ma}
\affiliation{Department of Physics, Beijing Normal University, Beijing 100875,  \textit{China}}
\email{txma@bnu.edu.cn}
\author{H.C. Kao}
\affiliation{Physics Department, National Taiwan Normal University, Taipei 11677, Taiwan, ROC}
\email{hckao@phy.ntnu.edu.tw}

\begin{abstract}
Mean field approach, although a generally reliable tool that captures major
short range correlations, often fails in symmetric low dimensional strongly
correlated electronic systems like those described by the Hubbard model. In
these situations a symmetry is ``almost broken". The problem is linked to
the restoration of the symmetry due to strong fluctuations (both quantum and
thermal) on all scales. The restoration of symmetry in statistical models of
scalar ``order parameter" fields was treated recently successfully on the
gaussian approximation level by symmetrization of the correlators. Here the
idea is extended to fermionic systems in which the order parameter is
composite. Furthermore the precision of the correlators can be improved
perturbatively. Such a scheme (based on covariant gaussian approximation) is
demonstrated on the 1D and 2D one band Hubbard models by comparison of the
correlator with exact diagonalization and MC simulations respectively.
\end{abstract}

\maketitle

%\pacs{74.20.Fg, 74.70.-b, 74.62.Fj}

\section{Introduction}

Thermal and quantum fluctuations play a much larger role in low dimensional
condensed matter systems than in three dimensional ones. As a consequence,
phase transitions to symmetry broken phases, exhibiting true long - range
order (LRO), like ferromagnet or antiferromagnet, are rare. In 2D only
systems possessing discrete symmetries can undergo finite temperature
spontaneous symmetry breaking, while in 1D they are ``forbidden" altogether.
The Mermin - Wagner theorem \cite{Mermin,Chaikin} states that fluctuations
for systems that have a continuous symmetry like the $SU\left( 2\right) $
symmetric Heisenberg model are strong enough to destroy LRO at any nonzero
temperature. The order parameter locally exists, but averages out due to
effective disordering of its ``phase" over the sample. To be specific, in
Heisenberg ferromagnet, the average of local order parameter, the spin
density, $\left \langle S^{i}\left( \mathbf{r}\right) \right \rangle =0$.

The symmetry therefore is not spontaneously broken in the low temperature
phase (that strictly speaking there is no ``symmetry breaking transition"
according to the Landau paradigm), yet the strong short range order (SRO) is
crucial for qualitative understanding of such systems ranging from high $%
T_{c}$ cuprate superconductors to quantum magnets. Despite vanishing
expectation value (VEV) order parameter, the correlator of the order
parameter, $P\left( \mathbf{r}\right) =\left \langle S^{i}\left( \mathbf{r}%
\right) S^{j}\left( \mathbf{0}\right) \right \rangle $, still characterizes
well the short range order. Generally it describes the spin excitations in
the system although there are no Goldstone bosons demanded by the continuous
symmetry breaking (via so called Ward identities). At least naively, the
symmetry is ``almost" broken in a sense that the correlator typically
decreases slowly (``local order" extends to large sizes). This contrasts
with that in true LRO phase in which the correlator approaches a constant at
large separation.

An approximate ``mean field" description of such systems having ``almost
long range order" very often results in various ``spurious" broken phases.
Within the Ginzburg - Landau - Wilson approach on\ the classical level,
phase diagrams contains host of ``broken symmetry" solutions. Very often it
is considered to be a failure of the approximation scheme, be it the
classical approximation, perturbation theory or a variational approach like
the mean field. One declares that the approximation is ``not capable" or
``fails to capture" the restoration of symmetry due to fluctuations and is
abandoned. Sometimes however an attempt was made to ``repair" such an
approximation by ``symmetrization" of the Green's functions (GF) calculated
starting with the symmetry broken solution.

In 2D statistical field theory of scalar fields the idea was attempted in
the framework of the ``shifted field" perturbation theory\cite{Jevicki}. It
worked well in models with discrete symmetries, but immediately ran into a
problem of with continuous symmetric SRO. Infrared divergencies appear at
low dimensions due to Goldstone modes. However it was shown that these
``spurious" divergencies generally cancel\cite{David}. In condensed matter
physics a similar problem was encountered in the context of thermal
fluctuations of the Abrikosov vortex lattice that appears in type II
superconductors in strong magnetic field. While calculating the spectrum of
thermal excitations of the 2D Abrikosov vortex lattice within the Ginzburg -
Landau theory, it was noticed\cite{Maki} that the gapless mode is softer
than the usual Goldstone mode expected as a result of spontaneous breaking
of translational invariance. At small k-vectors the correlator of the
superconducting order parameter field behaves as $1/k^{4}$. This unexpected
additional softness leads to infrared divergencies at higher orders. As a
result, the perturbation theory around the vortex state became doubtful
until it was realized that these divergencies are also spurious\cite{Kao}.
After the cancellation was established, symmetrization of the perturbative
GF are a way to get reasonable results for structure functions\cite{Lisolid}.

An interesting question is whether similar approach can be applied to
strongly coupled electronic systems directly on the microscopic level? The
symmetry breaking in such models (like the Hubbard, Heisenberg etc) is
necessarily ``dynamical" in a sense that the order parameter like the spin
density in a ferromagnet mentioned above is quadratic in the electron field
(not linear as in appears in the Ginzburg - Landau bosonic description).
Physically it means that there is a condensation of fermionic pairs
(excitons, Copper pairs...). Therefore generally these phases are not
approachable perturbatively and one has to either reexpress\ the theory it
terms of a bosonic field (bosonization) or use a nonperturbative method. The
simplest variational approach for which the (spurious) dynamical symmetry
breaking can be described is the gaussian (or Hartree - Fock) covariant
approximation described in detail for bosonic systems in ref.%
\onlinecite{Wang17} and fermionic many - body systems in ref.\onlinecite{CCA}%
.

In this paper we propose a ``symmetrization" method to study strongly
interacting electronic systems with strong LRO based on previous experience
with statistical physics expressed via order parameter directly\cite{Wang17}%
. It is tested on the benchmark models, the 1D and the 2D one band Hubbard
models for which exact diagonalization and Monte Carlo simulations are
performed. The symmetrization idea for ``almost broken" phases (sometimes
qualitatively described as ``preformed" correlated domains of the low
temperature phase or fluctuation dominated situations) is not new in physics.

The paper is organized as follows. In Section II the problem with standard
mean field type method in fermionic theories (known under various names in
different contexts as Hartree - Fock, BCS, exciton condensation...) is
presented. The solution to the problem in the strong SRO case by
symmetrization is proposed in Section III. Section IV contains its
application to the half filled or not half filled Hubbard model in $D=1,2$.
One can further improve the results expanding the self energy around the
gaussian solution (so called gaussian perturbation theory). This is done in
Section V. The results are compared with MC simulations in Section VI.
Results are summarized in Section VII.

\section{Spurious mean field symmetry breaking in fermionic models}

\subsection{Matsubara action for an interacting electron system}

Let us start with a general model of interacting fermions described by
Hamiltonian%
\begin{equation}
H=\sum \nolimits_{\mathbf{rr}^{\prime }}\left \{ -T_{\mathbf{rr}^{\prime
}}^{AB}a_{\mathbf{r}}^{A\dagger }a_{\mathbf{r}^{\prime }}^{B}+\frac{1}{2}V_{%
\mathbf{r-r}^{\prime }}a_{\mathbf{r}}^{A\dagger }a_{\mathbf{r}}^{A}a_{%
\mathbf{r}^{\prime }}^{B\dagger }a_{\mathbf{r}^{\prime }}^{B}\right \} \text{%
.}  \label{Hamiltonian}
\end{equation}%
where the band (valley) and spin denoted collectively by index $A$.
Summation over repeated indices is assumed. The hopping amplitudes $T_{%
\mathbf{rr}^{\prime }}$ typically extend to several nearest neighbours. The
interaction $V$ is assumed to be of the two - body (four Fermi) density -
density variety, appropriate to an effective description of many - body
electronic systems.

It is convenient for our purposes to describe it via path integral over a
large number of Grassmanian variables $\psi _{a}^{A}$. To simplify
notations, we initially lump position in space and Matsubara time into $%
a=\left \{ \mathbf{a\equiv r},a_{0}\equiv t\right \} $. Translation
invariance in $a$ is assumed. The Matsubara action corresponding to the
Hamiltonian therefore is:

\begin{equation}
\mathcal{A}=\psi _{a}^{\ast A}T_{a-b}^{AB}\psi _{b}^{B}+\frac{1}{2}\psi
_{a}^{A\ast }\psi _{a}^{A}V_{a-b}^{AB}\psi _{b}^{B\ast }\psi _{b}^{B}\text{.}
\label{action}
\end{equation}%
$V$ is symmetric under $A\leftrightarrow B,a\leftrightarrow b$. In modeling
strongly interacting systems in real space one typically considers hopping
on a lattice with periodic boundary conditions in each direction. For
simplicity we take a hypercubic lattice with lattice spacing defining the
unit of length and coordinates being integers $1,..N_{s}$, Matsubara time
(discretized as $t=1,...N_{t}$ with time step $\tau =\left( TN_{t}\right)
^{-1}$) on the segment from $0$ to $T^{-1}$, where $T$ is temperature. The
fermionic field is anti - periodic on the segment\cite{NO}.

Symmetry group $G$ (discrete or continuous) that might be spontaneously
broken consists of space - time independent (unitary) linear transformations
of the fermion field:%
\begin{equation}
\psi _{a}^{A}\rightarrow U^{AB}\psi _{a}^{B}\text{.}
\label{symmetrytransform}
\end{equation}%
As mentioned in Introduction, a general question arises. What happens when
fluctuations destroy the long range order, but an approximation incorrectly
``restores" the LRO? In fermionic system the fermionic field cannot have
nonzero expectation value, $\left \langle \psi _{a}^{A}\right \rangle =0$,
so to approach ``dynamical" SRO systems one can attempt to start with a
``mean field" variational solution of the order parameter quadratic in $\psi
_{a}^{A}$. An approximate Green's function, the expectation value,%
\begin{equation}
G_{a-b}^{AB}=\left \langle \psi _{a}^{\ast A}\psi _{b}^{B}\right \rangle ,
\label{corrdef}
\end{equation}%
is generally not invariant under the symmetry transformation,in the sense of
\begin{equation}
G_{a-b}^{AB}\neq U^{\ast AX}U^{BY}G_{a-b}^{XY}\text{.}  \label{symtransform}
\end{equation}

It is considered as a failure of the approximation scheme: the approximation
is \textquotedblleft not capable" or \textquotedblleft fails to capture" the
restoration of symmetry due to fluctuations. We try to take another shot at
these cases. The simplest variational approach for which the (spurious)
dynamical symmetry breaking can be described is the gaussian (or Hartree -
Fock) covariant approximation described in detail for bosonic systems in
ref. \onlinecite{Wang17} and fermionic many - body systems in ref. %
\onlinecite{CCA}.

\subsection{Gap equation and its symmetry broken solutions}

The HF variational GF is determined by the gap equation,

\begin{equation}
-\left[ G^{-1}\right] _{b-a}^{BA}=-T_{a-b}^{AB}-\delta _{a-b}\delta
^{AB}\sum \nolimits_{x,X}V_{x-a}^{XA}G_{0}^{XX}+V_{a-b}^{AB}G_{b-a}^{BA}%
\text{,}  \label{gapeq}
\end{equation}%
where the Green's function is a matrix with regard to indices $A$,$B$ and
$a$,$b$. In momentum space, defined by

\begin{equation}
\psi _{t,\mathbf{r}}^{A}=\sqrt{\frac{T}{N_{s}^{D}}}\sum%
\nolimits_{k_{1},...k_{D}=1}^{N_{s}}\sum \nolimits_{n=1}^{N_{t}}\exp \left[
i\left( \frac{\pi }{N_{t}}\left( 2n+1\right) t+\frac{2\pi }{N_{s}}\mathbf{%
k\cdot r}\right) \right] \psi _{n,\mathbf{k}}^{A}\text{,}
\label{Fourierfield}
\end{equation}%
the correlator is written as:

\begin{eqnarray}
G_{t-t^{\prime },\mathbf{r-r}^{\prime }}^{AB} &=&\frac{T}{N^{D}}\sum
\nolimits_{n\mathbf{k}}\exp \left[ -i\left( \frac{\pi }{N_{t}}\left(
2n+1\right) \left( t-t^{\prime }\right) +\frac{2\pi }{N_{s}}\mathbf{k\cdot }%
\left( \mathbf{r-r}^{\prime }\right) \right) \right] g_{n\mathbf{k}}^{AB}
\label{Fourier_corr} \\
G_{a,b}^{AB} &=&\frac{T}{N^{D}}\sum \nolimits_{\chi }\exp \left[ -i\left(
a-b\right) \cdot \chi \right] g_{\chi }^{AB}  \notag
\end{eqnarray}%
where in the last line, a shorthand space - time notations, $\sum
\nolimits_{\chi }\equiv \sum \nolimits_{nk}$ and $\chi \equiv \left \{ \frac{%
2\pi }{N_{t}}\left( n+1/2\right) ,\frac{2\pi }{N_{s}}\mathbf{k}\right \} $,$%
a=\left \{ t,\mathbf{r}\right \} $, $b=\left \{ t^{\prime },\mathbf{r}%
^{\prime }\right \} $ were used, and $a\cdot $ $\chi =t\times \frac{\pi }{%
N_{t}}\left( 2n+1\right) +\frac{2\pi }{N_{s}}\mathbf{k\cdot r}$. Similarly
it is convenient to define%
\begin{eqnarray}
T_{a-b}^{AB} &=&\frac{1}{TN_{t}^{2}N^{D}}\sum \nolimits_{\chi }\exp \left[
i\left( a-b\right) \cdot \chi \right] t_{\chi }^{BA};  \label{T_V_Fourier} \\
V_{a-b}^{AB} &=&\frac{1}{TN_{t}^{2}N^{D}}\sum \nolimits_{\chi }\exp \left[
i\left( a-b\right) \cdot \chi \right] v_{\chi }^{AB}\text{.}  \notag
\end{eqnarray}%
Consequently the Fourier transform of the gap equation reads:

\begin{equation}
\left[ g_{\omega }^{-1}\right] ^{BA}=-t_{\omega }^{BA}+\frac{T}{N_{s}^{D}}%
\sum \nolimits_{\chi }\left( v_{\omega -\chi }^{AB}g_{\chi }^{BA}-\delta
^{AB}v_{\lambda =0}^{AX}g_{\chi }^{XX}\right) \text{.}  \label{gapeq_Fourier}
\end{equation}%
where $\omega $ is also a shorthand space-time notation of the Fourier
indices like $\chi $. As an example let us consider the simplest example of
the \textquotedblleft quantum dot".

\subsection{Spurious magnetic phase of the quantum dot}

Let us consider the simplest Hamiltonian for a Pauli spinor $\psi ^{A}$,
with spin projections $A=\uparrow $ (up), $\downarrow $ (down):%
\begin{equation}
H=-\mu a^{A\dagger }a^{A}+Ua^{\uparrow \dagger }a^{\uparrow }a^{\downarrow
\dagger }a^{\downarrow }\text{.}  \label{QD_Hamiltonian}
\end{equation}%
This corresponds to the Matsubara action:

\begin{eqnarray}
A &=&\sum \nolimits_{t,s=1}^{N_{t}}\left \{ \psi _{t}^{A\ast
}T_{t-s}^{AB}\psi _{s}^{B}+\tau U\psi _{t}^{\upharpoonleft \ast }\psi
_{t}^{\upharpoonleft }\psi _{t}^{\downarrow \ast }\psi _{t}^{\downarrow
}\right \} \text{;}  \label{actionQD} \\
T_{t-s}^{AB} &=&\delta ^{AB}\left( \delta _{t+1-s}-\delta _{t-s}-\delta
_{ts}\tau \mu \right) \text{,}  \notag
\end{eqnarray}%
where $\tau =\left( TN_{t}\right) ^{-1}$. Comparing the interaction term to
that of the general action, Eq.(\ref{action}), one identifies:

\begin{equation}
V_{t-s}^{AB}=\tau U\delta _{t-s}\text{.}  \label{intHubbard}
\end{equation}

The time translation symmetry is fully utilized by using the Fourier
transforms,%
\begin{eqnarray}
t_{m}^{AB} &=&\delta ^{AB}\varepsilon _{m};\text{ }\varepsilon _{m}=\frac{1}{%
\tau }\left( \exp \left[ i\frac{2\pi }{N_{t}}\left( m+1/2\right) \right]
-1\right) -\mu \text{; }  \label{t_v} \\
\text{\  \ }v_{m}^{AB} &=&U\text{.}  \notag
\end{eqnarray}%
The gap equation takes a simple form,%
\begin{equation}
\left[ g_{m}^{-1}\right] ^{BA}=-t_{m}^{BA}+\Sigma ^{BA}\text{,}
\label{gapeq_FourierQD}
\end{equation}%
where the self energy,%
\begin{equation}
\Sigma ^{BA}=UT\sum \nolimits_{X,m}\left( g_{m}^{BA}-\delta
^{AB}g_{m}^{XX}\right) \text{,}  \label{sigma}
\end{equation}%
is frequency independent. The equation for the self energy thus becomes
algebraic :%
\begin{equation}
\Sigma ^{AB}=U\left( \delta ^{AB}\sum \nolimits_{X}n^{XX}-n^{AB}\right)
\text{.}  \label{sigmaQD2}
\end{equation}%
The four density components, $n^{AB}=T\sum \nolimits_{m}g_{m}^{AB}$, are
variational parameters. We can narrow the search, if the residual $U\left(
1\right) $ symmetry of spin rotations around the $z$ axis is assumed (of
course any other direction can be chosen). This ensures that $n^{\uparrow
\downarrow }=n^{\downarrow \uparrow }=0$, and only two parameters are left, $%
n^{\uparrow \uparrow }=n^{\uparrow }$ and $n^{\downarrow \downarrow
}=n^{\downarrow }$. Therefore one gets two equations%
\begin{equation}
g_{m}^{AA}=-\frac{1}{\varepsilon _{m}+Un^{\overline{A}}}\text{,}
\label{gdef1}
\end{equation}%
where the bar means the spin $A$ reversal: $\overline{\uparrow }=\downarrow $
and $\overline{\downarrow }=\uparrow $.

The gap equation in terms of densities subsequently becomes algebraic:%
\begin{equation}
n^{A}=-T\sum \nolimits_{m}\frac{1}{\varepsilon _{m}+Un^{\overline{A}}}%
=f\left( n^{\overline{A}}\right) \text{.}  \label{neq}
\end{equation}%
The last lines are the case of infinite $N_{t}$ in which%
\begin{eqnarray}
\varepsilon _{m} &=&i\omega _{m}-\mu ;  \label{epsilon} \\
\omega _{m} &=&\pi T\left( 2m+1\right) ,  \notag
\end{eqnarray}%
and the summation results in the Fermi-Dirac distribution%
\begin{equation}
f\left( n^{A}\right) \equiv \frac{1}{\exp \left[ \left( Un^{A}-\mu \right) /T%
\right] +1}\text{.}  \label{Fermi}
\end{equation}%
The nonmagnetic solution, $n^{\uparrow }=n^{\downarrow }$, is trivial at
half filling, for which the electron-hole symmetry ensures $n^{\uparrow
}+n^{\downarrow }=1$, $\mu =U/2$, $n^{\uparrow }=1/2$. As a result the HF GF
is independent of coupling $U$:
\begin{equation}
g_{m}^{AB}=\delta ^{AB}\frac{i}{\omega _{m}}\text{.}  \label{para}
\end{equation}%
This (imaginary part is the horizontal green segment in Fig. 1) deviates
significantly from the exact value represented by the red line.

The model at half filling has just one parameter $u\equiv U/T$. The magnetic
solution with magnetization, $M=\frac{1}{2}\left( n^{\uparrow
}-n^{\downarrow }\right) =n^{\uparrow }-1/2$, of the gap equation,
\begin{equation}
\exp \left[ uM\right] =\frac{1/2+M}{1/2-M}\text{,}  \label{ferroeq}
\end{equation}%
exists above the spurious second order transition point, $u_{c}=4$. We will
use this toy model to exemplify the symmetrization idea in the following
Section.

\section{Symmetrized Green's functions approach}

\subsection{Qualitative description of the symmetrization}

It was shown\cite{Wang17} for the case of bosonic low dimensional models
that in the strong coupling regime, where within classical or gaussian
approximation the symmetry is ``spuriously broken", the symmetrized
nonsymmetric Green's functions is quite close to exact or Monte Carlo
calculated result. It means that symmetrization of the GF effectively takes
into account highly correlated domains. Of course a more rigorous approach
would divide the degrees of freedom into two scales, large distance
correlations, LRO and short distance correlations, SRO. It can be performed
for certain bosonic models using renormalization group ideas, especially
when the Berezinskii - Kosterlitz - Thouless transition is involved. However
such an approach is extremely complicated in fermionic models in which order
parameter is quadratic in fermionic operators (condensation of pairs). The
simplistic symmetrization approach that does not involve the explicit
separation of scales, however is still effective, as we demonstrate in
following Sections. The symmetrization qualitatively takes into account the
largest available scale by ``averaging over" the global symmetry group.

Here we generalize the approach to a general interacting fermionic model in
which (on the mean field level) the global (space and time independent)
symmetry group $\mathcal{G}$ is spontaneously broken down to its subgroup $%
\mathcal{H}$. The half filled quantum dot of the previous Section can serve
as a toy model in which for $U>U_{c}=4T$ the symmetry group $\mathcal{G}%
=SU\left( 2\right) $ (all the spin rotations, Eq.(\ref{symtransform}))
is``spontaneously broken" to its subgroup $\mathcal{H}=U\left( 1\right) $
(rotations around an axis determined by the breaking direction, in our case
the $z$ axis).

\subsection{Formulation of the symmetrization approach}

Generally an approximate GF is symmetrized using the so called invariant
Haar measure over the group $\mathcal{G}$\cite{Haarmeasure}:%
\begin{eqnarray}
\left \langle \psi _{a_{1}}^{\ast A_{1}}...\psi _{a_{n}}^{\ast A_{n}}\psi
_{b_{1}}^{B_{1}}...\psi _{b_{n}}^{B_{n}}\right \rangle _{sym} &=&\int dU%
\text{ }U^{\ast A_{1}X_{1}}...U^{\ast
A_{n}X_{n}}U^{B_{1}Y_{1}}...U^{B_{n}Y_{n}}  \label{general_sym} \\
&&\text{ \  \  \  \  \  \  \  \ }\times \left \langle \psi _{x_{1}}^{\ast
X_{1}}...\psi _{x_{n}}^{\ast X_{n}}\psi _{y_{1}}^{Y_{1}}...\psi
_{y_{n}}^{Y_{n}}\right \rangle \text{.}  \notag
\end{eqnarray}%
The mathematical definition of the measure $dU$ for compact Lie groups is
available in literature where it is shown that it is unique. We provide here
simple examples starting from $\mathcal{G}=U\left( 1\right) $. In this case,
the group elements are described by a 2D rotation angle $\theta $, and Haar
measure is just angle average, $\int dU$ $f\left[ U\right] =\frac{1}{2\pi }%
\int_{\theta =0}^{2\pi }d\theta f\left[ \theta \right] $. In our case $%
\mathcal{G}=SU\left( 2\right) $, the integration over the group reduces to
the following integral over three Euler angles parameterizing rotations of
the spin\cite{Haarmeasure}: $\int dU$ $f\left[ U\right] =\frac{1}{\left(
2\pi \right) ^{2}}\int_{0}^{\pi }d\psi \sin \psi \int_{0}^{\pi }d\theta \sin
^{2}\theta \int_{0}^{2\pi }d\varphi $ $f\left[ \psi ,\theta ,\varphi \right]
$. Actually the integration over the vacuum manifold $\mathcal{G}/\mathcal{H}
$ only (just two angles) is sufficient for most applications. For discrete
groups the symmetrization becomes a rather obvious summation over all the
group elements.

We will need only the following basic $\mathcal{G=}SU\left( N\right) $
integrals\cite{Rossi}, for the fundamental representation%
\begin{equation}
\int U^{\ast AX}U^{BY}dU=\frac{1}{N}\delta ^{AB}\delta ^{XY}\text{,}
\label{SU2}
\end{equation}%
and
\begin{eqnarray}
&&\int U^{\ast A_{1}X_{1}}U^{\ast A_{2}X_{2}}U^{B_{1}Y_{1}}U^{B_{2}Y_{2}}dU
\label{4U} \\
&=&\frac{1}{N^{2}-1}\left \{
\begin{array}{c}
\delta ^{A_{1}B_{1}}\delta ^{A_{2}B_{2}}\delta ^{X_{1}Y_{1}}\delta
^{X_{2}Y_{2}}+\delta ^{A_{1}B_{2}}\delta ^{A_{2}B_{1}}\delta
^{X_{1}Y_{2}}\delta ^{X_{2}Y_{1}} \\
-\frac{1}{N}\left( \delta ^{A_{1}B_{2}}\delta ^{A_{2}B_{1}}\delta
^{X_{1}Y_{1}}\delta ^{X_{2}Y_{2}}+\delta ^{A_{1}B_{1}}\delta
^{A_{2}B_{2}}\delta ^{X_{1}Y_{2}}\delta ^{X_{2}Y_{1}}\right)%
\end{array}%
\right \} \text{.}  \notag
\end{eqnarray}%
As an example, let us symmetrize the one - body electron and the spin
correlator that is a two - body correlator in the single band Hubbard model.
The symmetrized correlator reads:%
\begin{eqnarray}
\left \langle \psi _{a}^{\ast A}\psi _{b}^{B}\right \rangle _{sym} &=&\int
U^{\ast AX}U^{BY}dU\left \langle \psi _{a}^{\ast X}\psi _{b}^{Y}\right
\rangle  \label{twobody} \\
&=&\frac{1}{2}\delta ^{AB}\delta ^{XY}\left \langle \psi _{a}^{\ast X}\psi
_{b}^{Y}\right \rangle =\frac{1}{2}\delta ^{AB}\left( \left \langle \psi
_{a}^{\ast \uparrow }\psi _{b}^{\uparrow }\right \rangle +\left \langle \psi
_{a}^{\ast \downarrow }\psi _{b}^{\downarrow }\right \rangle \right) \text{.}
\notag
\end{eqnarray}

The spin correlator has the following symmetrized form%
\begin{eqnarray}
\left \langle S_{a}^{i}S_{b}^{j}\right \rangle _{sym} &=&\frac{1}{4}\left
\langle \psi _{a}^{\ast A_{1}}\sigma _{i}^{A_{1}B_{1}}\psi _{a}^{B_{1}}\psi
_{b}^{\ast A_{2}}\sigma _{j}^{A_{2}B_{2}}\psi _{b}^{B_{2}}\right \rangle
_{sym}  \label{spincorr} \\
&=&\frac{1}{4}\sigma _{i}^{A_{1}B_{1}}\sigma _{j}^{A_{2}B_{2}}\int U^{\ast
A_{1}X_{1}}U^{B_{1}Y_{1}}U^{\ast A_{2}X_{2}}U^{B_{2}Y_{2}}dU\left \langle
\psi _{a}^{\ast X_{1}}\psi _{a}^{Y_{1}}\psi _{b}^{\ast X_{2}}\psi
_{b}^{Y_{2}}\right \rangle \text{.}  \notag
\end{eqnarray}%
Using the group integral of Eq.(\ref{4U}), one obtains%
\begin{eqnarray}
\left \langle S_{a}^{i}S_{b}^{j}\right \rangle _{sym} &=&\frac{1}{12}\sigma
_{i}^{AB}\sigma _{j}^{BA}\left( \left \langle \psi _{a}^{\ast X}\psi
_{a}^{Y}\psi _{b}^{\ast Y}\psi _{b}^{X}\right \rangle -\frac{1}{2}\left
\langle \psi _{a}^{\ast X}\psi _{a}^{X}\psi _{b}^{\ast Y}\psi _{b}^{Y}\right
\rangle \right)  \label{spincorr2} \\
&=&\frac{1}{6}\delta ^{ij}\left( \left \langle \psi _{a}^{\ast X}\psi
_{a}^{Y}\psi _{b}^{\ast Y}\psi _{b}^{X}\right \rangle -\frac{1}{2}\left
\langle \psi _{a}^{\ast X}\psi _{a}^{X}\psi _{b}^{\ast Y}\psi _{b}^{Y}\right
\rangle \right) \text{.}  \notag
\end{eqnarray}%
The density correlator on the other hand is already symmetrized:

\begin{eqnarray}
\left \langle n_{a}n_{b}\right \rangle &=&\left \langle \psi _{a}^{\ast
A}\psi _{a}^{A}\psi _{b}^{\ast B}\psi _{b}^{B}\right \rangle _{sym}
\label{denden} \\
&=&\int U^{\ast AX_{1}}U^{AY_{1}}U^{\ast BX_{2}}U^{BY_{2}}dU\left \langle
\psi _{a}^{\ast X_{1}}\psi _{a}^{Y_{1}}\psi _{b}^{\ast X_{2}}\psi
_{b}^{Y_{2}}\right \rangle =\left \langle \psi _{a}^{\ast X}\psi
_{a}^{X}\psi _{b}^{\ast Y}\psi _{b}^{Y}\right \rangle \text{.}  \notag
\end{eqnarray}%
Sometimes this is expressed in the Wigner - Eckart form that only symmetric
quantities like $\left \langle \psi ^{\ast \uparrow }\psi ^{\uparrow
}\right
\rangle +\left \langle \psi ^{\ast \downarrow }\psi ^{\downarrow
}\right
\rangle $ the ones that can be calculated using the symmetrization
approach\cite{David}.

Before applying the procedure to the Hubbard model, let us exemplify
advantages of the approach on the simplest fermionic toy model in $D=0$,
where the symmetry restoration phenomenon is expected to be the strongest.

\subsection{The toy model example}

The quantum dot at half filling of the previous Section can be exactly solved%
\cite{CCA}. The correlator (all the energies like the coupling $U$ are in
units of $T$):

\begin{equation}
g_{m}=\frac{i\pi \left( 2m+1\right) }{\pi ^{2}\left( 2m+1\right) ^{2}+u^{2}/4%
}\text{,}  \label{QDexact}
\end{equation}%
where Matsubara frequency is $2m+1$($T=1$ now). The symmetric (paramagnetic)
solution result of Eq.(\ref{para}), $g_{m}=\frac{i}{\pi \left( 2m+1\right) }$%
, is independent of $u$ and thus pretty bad everywhere but close to $u=0$.
The paramagnetic (green line) and the exact (red line) correlators are given
as functions of $u$ in Fig.1 for $m=0,1$ (that is for Matsubara frequencies,
$\pi T$ and $3\pi T$ on Fig.1a and b) respectively. The correlator grossly
overestimates the exact one at the spurious critical point $u_{c}=4$, marked
in Fig.1 by a dashed black line.

The magnetic solution of Eq.(\ref{ferroeq}), symmetrized according to Eq.(%
\ref{twobody}) above, takes a form:
\begin{eqnarray}
g_{m}^{AB} &=&\delta ^{AB}g_{m};  \label{symCGA} \\
g_{m} &=&\frac{i\pi \left( 2m+1\right) }{\pi ^{2}\left( 2m+1\right)
^{2}+u^{2}\left( n^{\downarrow }-1/2\right) ^{2}}\text{.}  \notag
\end{eqnarray}%
The value of density $n^{\downarrow }$ here was calculated numerically by
solving Eq.(\ref{ferroeq}). It is given in Fig.1 as the dark green line. One
observes that, while the large $u$ asymptotics is exact, at intermediate
couplings the agreement is on the 10\% level. The perturbative correction is
also presented and in section IV we will discuss how one can perturbatively
improve the approximation (perturbative correction leading to the result
represented by the violet line). The \textquotedblleft almost" broken phase
symmetrized HF, Eq.(\ref{symCGA}) becomes asymptotically correct at large
couplings. As Fig.1b demonstrates, for higher Matsubara frequencies the
approximation very fast becomes excellent in the whole range of parameters.
Of course the large $m$ asymptotics is guaranteed.

Now we apply this method to more complicated solvable models of strongly
interacting electron systems. The prime example is the one band Hubbard
model that describes qualitatively well several manufactured 2D quantum
magnets and 1D and 2D BEC systems.

\begin{figure}[tbp]
\begin{center}
\includegraphics[width=12cm]{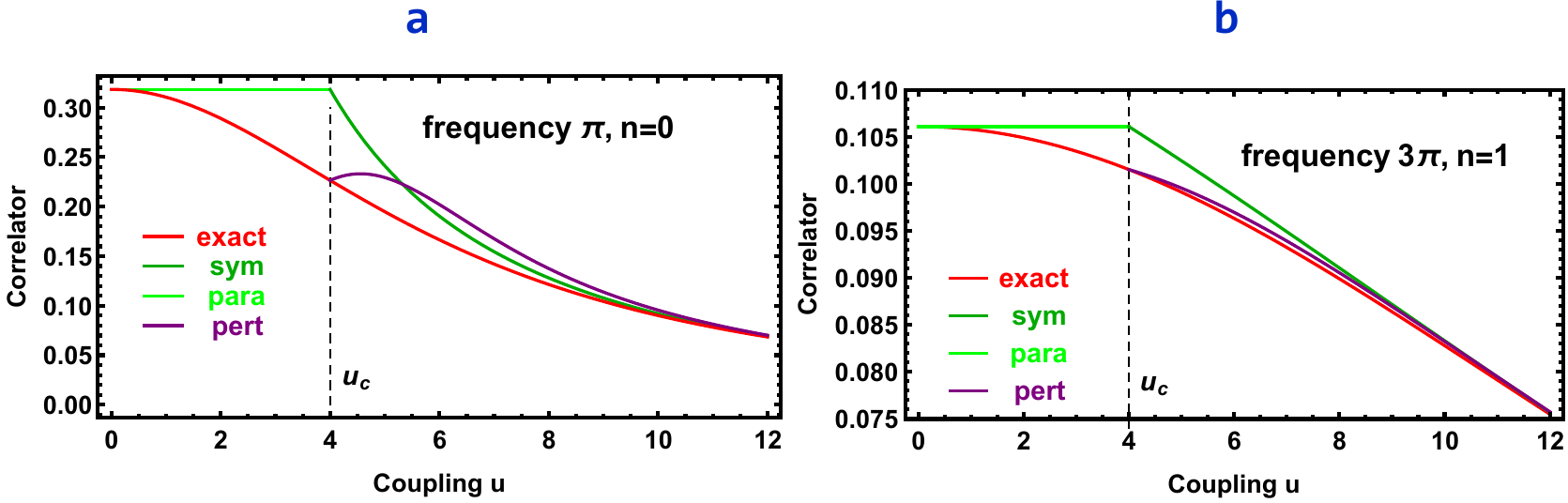}
\end{center}
\par
\vspace{-0.5cm}
\caption{ Imaginary part of the correlator for quantum dot at half filling
in wide range of couplings u=U/T. Matsubara frequency is $\protect \omega %
_{n}=\protect \pi T\left( 2n+1\right) =\protect \pi $, at $n=0,$ $T=1$. The
red line is the exact result, the green line is the Hartree Fock result, the
darker green line is the symmetrized green function Eq.(\protect \ref{symCGA})
for the magnetic phase,\ and the purple line is the perturbative correction
to gaussian approximation (PCGA) Eqs.(\protect \ref{pgQD}),(\protect \ref{PCGA_QD}).}
\end{figure}

\section{Application to the antiferromagnetic SRO in the half filled Hubbard
model}

\subsection{ The Hubbard model.}

The single band Hubbard model for strongly interacting electrons is defined
on $D$ dimensional hypercubic lattice compactified in all directions into a
circle of perimeter $N_{s}$. The tunneling amplitude to the neighbouring
site is denoted in literature by $t$. We chose it to be the unit of energy $%
t=1$. Similarly the lattice spacing sets the unit of length $a=1$ and $\hbar
=1$. The Hamiltonian is (restricting for notational simplicity to one band
and $D=1$, although generalization to arbitrary $D$ and other types of
lattices is straightforward):

\begin{equation}
H=\sum \nolimits_{x=1}^{N_{s}}\left \{ -\left( a_{x}^{\alpha \dagger
}a_{x+1}^{\alpha }+h.c.\right) -\mu n_{x}+Un_{x}^{\upharpoonleft
}n_{x}^{\downarrow }\right \} \text{.}  \label{HubbardH}
\end{equation}%
The chemical potential $\mu $ and the on - site repulsion energy $U$ are
therefore given in units of the hopping energy. The spin index takes two
values $\alpha =\uparrow ,\downarrow $. The density and its spin components
are $n_{x}=n_{x}^{\upharpoonleft }+n_{x}^{\downarrow }$ with $n_{x}^{\alpha
}\equiv a_{x}^{\alpha \dagger }a_{x}^{\alpha }$ respectively. It is well
known that at half filling $\mu =\frac{U}{2}$ due to the particle - hole
symmetry\cite{Korepin}. Approximations we will use are ``covariant"\cite{CCA}
and thus respect this restriction.

The discretized Matsubara action is\cite{NO},
\begin{equation}
\mathcal{A}=\tau \sum \nolimits_{t,x}\frac{1}{\tau }\left( \psi
_{t+1,x}^{\alpha \ast }\psi _{t,x}^{\alpha }-\psi _{t,x}^{\alpha \ast }\psi
_{t,x}^{\alpha }\right) \ -\frac{1}{2}\left( \psi _{t,x}^{\alpha \ast }\psi
_{t,x+1}^{\alpha }+\psi _{t,x}^{\alpha \ast }\psi _{t,x-1}^{\alpha }\right)
-\mu n_{x}-U\psi _{t,x}^{\upharpoonleft \ast }\psi _{t,x}^{\downarrow \ast
}\psi _{t,x}^{\upharpoonleft }\psi _{t,x}^{\downarrow }\text{,}
\label{Liuaction}
\end{equation}%
where $n_{t,r}\equiv \psi _{t,x}^{\sigma \ast }\psi _{t,x}^{\sigma }$.
Generally for $D\leq 2$ and the nonabelian symmetry group symmetry
fluctuations (quantum and thermal) destroy numerous \textquotedblleft mean
field broken" phases, although previously attempted variational approaches
like the CGA at large coupling start from a \textquotedblleft broken" phase
solution of the minimization equations sometimes give a much better result
upon symmetrization. The start from recounting the well known HF gap
equation and its paramagnetic solution\cite{Auerbach}.

\subsection{The paramagnetic Hartree - Fock solution}

The hopping matrix and interaction in frequency-momentum space of the
corresponding Matsubara action, is:
\begin{eqnarray}
t_{n,k}^{\alpha \beta } &=&\delta ^{\alpha \beta }t_{n,k};\text{ \ }%
t_{n,k}=\varepsilon _{m}-2\cos \left[ \frac{2\pi }{N_{s}}k\right] \text{,}
\label{t_v_1D} \\
v_{n,k}^{\alpha \beta } &=&U\text{.}  \notag
\end{eqnarray}%
The gap equation in paramagnet simplifies to

\begin{equation}
\Sigma =-U\frac{T}{N_{s}}\sum \nolimits_{m,k}g_{mk}\text{,}
\label{gapeq_para}
\end{equation}%
and is solved numerically (for infinite $N_{t}$) for $T=0.2$. For half
filling, $\mu =\frac{U}{2}$, the solution of the above equation is $\Sigma =%
\frac{U}{2}$. The results for the imaginary part of the correlator $g_{mk}$
presented for $N_{s}=4$ in Fig. 2 as the green line for couplings not
exceeding the spurious critical value of $U_{c}\approx 1.3335$. Frequency is
the lowest, $n=0$ corresponding to $\omega =\pi T$, while quasimomentum $k=0$
in Fig. 2 and $k=N_{s}/4=1$ ($k$ - vector $\pi /\left( 2a\right) $ in
physical units) in Fig. 3 but post gaussian correction, or the perturbative
correction to gaussian approximation, PCGA is good (PCGA theory will be
presented in Sec. VA). As for the quantum dot, it (case for $N_{s}=4$ ) also
does not compare well with the exact diagonalization result (red line) for
coupling that is not very small. The real part of the correlator on the
Fermi surface for $k=\pi /2$ is zero.

Similar results are obtained for other physical quantities at $D=1$, while
generalization to $2D$ gives results presented in Fig. 8 that will be
commented below. The problem for $U<U_{c}$ is easily remedied by a
perturbative correction described in Section IV.

\begin{figure}[tbp]
\begin{center}
\includegraphics[width=10cm]{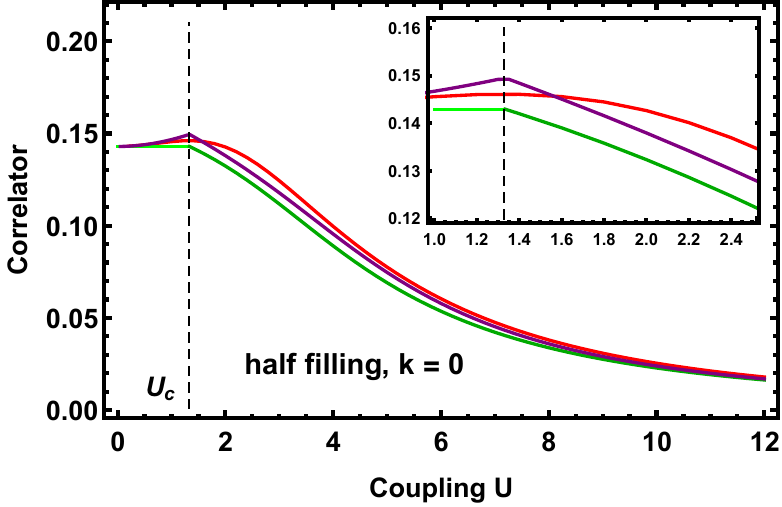}
\end{center}
\par
\vspace{-0.5cm}
\caption{Comparison of the exact correlator of a short Hubbard chain at
temperature $T=0.2$ with approximations in wide range of couplings for $%
\protect \omega =\protect \pi T$, $k=0$. The approximations include the CGA
(green lines, the para solution from Eq.(\protect \ref{gapeq_para}), darker
green lines from Eq.(\protect \ref{AF_sym})) that is symmetrized above the
spurious transition at $U_{c}=1.33$, and perturbative correction to gaussian
approximation (PCGA, purple lines).}
\end{figure}

\subsection{Symmetrized anti - ferromagnetic Green's function}

\subsubsection{Spin rotation and translation spurious symmetry breaking on
the HF level}

The spin $SU\left( 2\right) $ symmetry of the Hubbard model at half filling
and large $U$ is spontaneously broken on the HF level to its $U\left(
1\right) $ subgroup chosen here as rotation around the $z$ spin direction.
Simultaneously the translation symmetry is broken, so that two sublattices $%
I=1,2$ appear. Therefore translational symmetry becomes smaller with unit
cell index $x^{\prime }=1,...,N^{\prime }$ with $N^{\prime }=N_{s}/2$. The
position for the sublattice $1$ is $x=2x^{\prime }-1$, while for sublattice $%
2$ becomes $x=2x^{\prime }$. The Matsubara action therefore is rearranged as
a \textquotedblleft folded" one:
\begin{eqnarray}
A &=&\tau \sum \nolimits_{t,x^{\prime }}\frac{1}{\tau }\left( \psi
_{t-1,x^{\prime }}^{I\sigma \ast }\psi _{t,x^{\prime }}^{I\sigma }-\psi
_{t,x^{\prime }}^{I\sigma \ast }\psi _{t,x^{\prime }}^{I\sigma }\right)
-\psi _{t,x^{\prime }}^{I\sigma \ast }\sigma _{x}^{IJ}\psi _{t,x^{\prime
}}^{J\sigma }-\frac{1}{2}\psi _{t,x^{\prime }}^{I\sigma \ast }\left( \sigma
_{x}^{IJ}+i\sigma _{y}^{IJ}\right) \psi _{t,x^{\prime }-1}^{J\sigma }
\label{folded_action} \\
&&-\frac{1}{2}\psi _{t,x^{\prime }}^{I\sigma \ast }\left( \sigma
_{x}^{IJ}-i\sigma _{y}^{IJ}\right) \psi _{t,x^{\prime }+1}^{J\sigma }-\frac{U%
}{2}n_{i}^{I}+U\psi _{t,x^{\prime }}^{I\upharpoonleft \ast }\psi
_{t,x^{\prime }}^{I\downarrow \ast }\psi _{t,x^{\prime }}^{I\upharpoonleft
}\psi _{t,x^{\prime }}^{I\downarrow }  \notag
\end{eqnarray}%
Here summation over sublattice indices $I,J$ is assumed. The Fourier
transform now takes a form%
\begin{equation}
\psi _{it}^{I\sigma \ast }=\sqrt{\frac{T}{N^{\prime }}}\sum
\nolimits_{k^{\prime }=1}^{N^{\prime }}\sum \nolimits_{n=1}^{N_{t}}\exp %
\left[ i\left( -\frac{2\pi k^{\prime }}{N^{\prime }}i-\frac{2\pi \left(
n+1/2\right) }{N_{t}}t\right) \right] \psi _{k^{\prime }n}^{I\sigma \ast }
\label{FT_AF}
\end{equation}%
folded integer quasimomentum $k^{\prime }=1,...,N^{\prime }$. The action
becomes that of Eq.(\ref{action}) with%
\begin{eqnarray}
t_{nk^{\prime }}^{-1} &=&\delta ^{IJ}\varepsilon _{n}-\left( 1+\cos \left[
\frac{2\pi }{N^{\prime }}k^{\prime }\right] \right) \sigma _{x}^{IJ}-\sin %
\left[ \frac{2\pi }{N^{\prime }}k^{\prime }\right] \sigma _{y}^{IJ};
\label{AF_tv} \\
v_{nk^{\prime }}^{JI\alpha \beta } &=&U\delta ^{IJ},  \notag
\end{eqnarray}%
where $\varepsilon _{n}$ was defined in Eq.(\ref{t_v}).

The gap equation, Eq.(\ref{gapeq}), now take the following form:%
\begin{equation}
\Sigma ^{IJ\alpha \beta }=\frac{TU}{N^{\prime }}\delta ^{IJ}\sum
\nolimits_{nk^{\prime }}\left( \delta ^{\alpha \beta }g_{nk^{\prime
}}^{II\kappa \kappa }-g_{nk^{\prime }}^{II\beta \alpha }\right) =U\delta
^{IJ}\left( \delta ^{\alpha \beta }n^{II\kappa \kappa }-n^{II\beta \alpha
}\right) \text{.}  \label{gapeq_1D}
\end{equation}%
As is well known, it is solved by the anti - ferromagnetic (AF) Ansatz%
\begin{eqnarray}
n^{11\uparrow \uparrow } &=&n^{22\downarrow \downarrow
}=n_{1};n^{11\downarrow \downarrow }=n^{22\uparrow \uparrow }=n_{2};
\label{AF_Ansatz} \\
n^{11\uparrow \downarrow } &=&n^{11\downarrow \uparrow }=n^{22\uparrow
\downarrow }=n^{22\downarrow \uparrow }=0\text{.}  \notag
\end{eqnarray}%
The resulting algebraic equations at infinite $N_{t}$ are $n_{1}+n_{2}=1$,
and, defining magnetization, $M=n_{1}-\frac{1}{2}$,

\begin{gather}
\sum \nolimits_{k^{\prime }=1}^{N^{\prime }}\frac{1}{e_{k^{\prime }}}\tanh %
\left[ \frac{e_{k^{\prime }}}{2T}\right] =\frac{N_{s}}{U};
\label{algebraic_gapeq} \\
e_{k^{\prime }}^{2}=4\cos ^{2}\left[ \frac{2\pi }{N_{s}}k^{\prime }\right]
+\left( UM\right) ^{2}\text{.}  \notag
\end{gather}%
The spurious critical coupling therefore is:
\begin{equation}
U_{c}\left( T\right) =2N_{s}\left \{ \sum \nolimits_{k^{\prime
}=1}^{N^{\prime }}\cos ^{-1}\left[ \frac{2\pi }{N_{s}}k^{\prime }\right]
\tanh \left[ \frac{2\cos \left[ \frac{2\pi }{N_{s}}k^{\prime }\right] }{2T}%
\right] \right \} ^{-1}\text{.}  \label{Uc}
\end{equation}%
For particular cases shown in Figs. 2,3 and 5, we set $N_{s}=4$ and $24$
respectively. The values of the critical coupling at temperature $T=0.2$ are
$U_{c}=1.3335$ and $U_{c}=2.017$ at half filling respectively.

The nonsymmetrized correlator is diagonal in spin, $g_{nk^{\prime
}}^{IJ\uparrow \downarrow }=$ $g_{nk^{\prime }}^{IJ\downarrow \uparrow }=0$,
due to the residual $U\left( 1\right) $ symmetry, so we specify the spin $%
\alpha $ once:

\begin{equation}
g_{nk^{\prime }}^{IJ\alpha }=\frac{1}{e_{k^{\prime }}^{2}+\omega _{n}^{2}}%
\left \{ \left( 1+\cos \left[ \frac{2\pi }{N^{\prime }}k^{\prime }\right]
\right) \sigma _{x}^{IJ}-\sin \left[ \frac{2\pi }{N^{\prime }}k^{\prime }%
\right] \sigma _{y}^{IJ}+i\omega _{n}\delta ^{IJ}-sgn\left[ \alpha \right]
UM\sigma _{z}^{IJ}\right \} \text{.}  \label{gAF}
\end{equation}%
Here $sgn\left[ \alpha \right] =\sigma _{z}^{\alpha \alpha }$, namely $+1$
for $\uparrow $ and $-1$ for $\downarrow $. Recall that we have chosen the
direction of magnetization at large coupling in the spuriously broken phase
to be parallel to the $z$ spin direction. This should be symmetrized over
all the AF ground states.

\begin{figure}[tbp]
\begin{center}
\includegraphics[width=18cm]{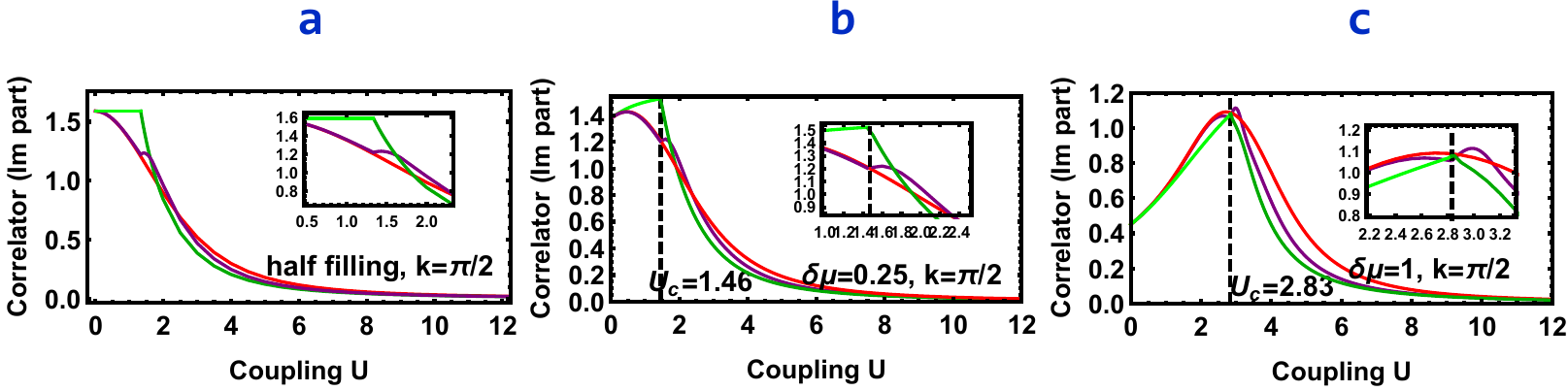}
\end{center}
\par
\vspace{-0.5cm}
\caption{Comparison of the exact correlator of a short Hubbard chain at
temperature $T=0.2$ with approximations in wide range of couplings for $%
\protect \omega =\protect \pi T$, $k=\protect \pi /2$ at $\protect \delta
\protect \mu =0,0.25,1$ from top to bottom. The approximations include the
CGA (green lines, the para solution from Eq.(\protect \ref{gapeq_para}),
darker green lines from Eq.(\protect \ref{AF_sym})) that is symmetrized above
the spurious transition at $U_{c}=1.33$, and perturbative correction to
gaussian approximation in Fig.3a (PCGA, purple lines). The inset of Fig.3a is
an enlarged figure near the ``critical coupling" region. Fig.3b and Fig.3c ($%
\protect \delta \protect \mu =0.25,1$ respectively), the results for the CGA
(green lines for para solution, darker green lines from the symmetrized
correlator) are present, the exact results are plotted as red lines.}
\end{figure}

\subsubsection{Symmetrization}

Then symmetry breaking pattern for the $SU\left( 2\right) \rightarrow
U\left( 1\right) $ for the paramagnet to AF involves simultaneous
translation symmetry breaking resulting in sublattices. Taking trace over
spins and dividing by $2$, the symmetrized frequency - quasimomentum GF is:$%
\left \langle \psi _{nx}^{\ast \uparrow }\psi _{ny}^{\uparrow
}\right
\rangle \approx \frac{1}{2}\left \langle \psi _{nx}^{\ast \sigma
}\psi _{ny}^{\sigma }\right \rangle _{AF}=\frac{1}{2}\left( \left \langle
\psi _{nx}^{\ast \uparrow }\psi _{ny}^{\uparrow }\right \rangle
_{AF}+\left
\langle \psi _{nx}^{\ast \downarrow }\psi _{ny}^{\downarrow
}\right \rangle _{AF}\right) $ ($n$ is the Matsubara frequency, $x,y$ are
the lattice coordinates)
\begin{eqnarray}
g_{nk}^{sym} &=&\frac{1}{N_{s}}\sum \nolimits_{x,y=1}^{N_{s}}\exp \left[
\frac{2i\pi k}{N_{s}}\left( x-y\right) \right] \left \langle \psi
_{nx}^{\ast \uparrow }\psi _{ny}^{\uparrow }\right \rangle  \label{symdef} \\
&\approx &\frac{1}{2N_{s}}\sum \nolimits_{x,y=1}^{N_{s}}\exp \left[ \frac{%
2i\pi k}{N_{s}}\left( x-y\right) \right] \left \langle \psi _{nx}^{\sigma
\ast }\psi _{ny}^{a\sigma }\right \rangle _{AF}\text{.}
\end{eqnarray}%
In sublattice notations this becomes:$\left \langle \psi _{nx}^{\ast
\uparrow }\psi _{ny}^{\uparrow }\right \rangle \approx \frac{1}{2}\left(
\left \langle \psi _{nx}^{\ast \uparrow }\psi _{ny}^{\uparrow
}\right
\rangle _{AF}+\left \langle \psi _{nx+1}^{\ast \uparrow }\psi
_{ny+1}^{\uparrow }\right \rangle _{AF}\right) $
\begin{eqnarray}
g_{nk}^{sym} &=&\frac{1}{4N^{\prime }}\sum \nolimits_{x^{\prime }y^{\prime
}}^{N^{\prime }}\exp \left[ \frac{2i\pi k}{N^{\prime }}\left( x^{\prime
}-y^{\prime }\right) \right] \left \{
\begin{array}{c}
\left \langle \psi _{nx^{\prime }}^{1\sigma \ast }\psi _{ny^{\prime
}}^{1\sigma }\right \rangle +\exp \left[ -\frac{i\pi k}{N^{\prime }}\right]
\left \langle \psi _{nx^{\prime }}^{1\sigma \ast }\psi _{ny^{\prime
}}^{2\sigma }\right \rangle \\
+\exp \left[ \frac{i\pi k}{N^{\prime }}\right] \left \langle \psi
_{nx^{\prime }}^{2\sigma \ast }\psi _{ny^{\prime }}^{1\sigma }\right \rangle
+\left \langle \psi _{nx^{\prime }}^{2\sigma \ast }\psi _{ny^{\prime
}}^{2\sigma }\right \rangle%
\end{array}%
\right \}  \label{sym_der} \\
&=&\frac{1}{4}\left \{
\begin{array}{c}
\left \langle \psi _{n,\text{mod}\left[ k\right] }^{1\sigma \ast }\psi _{n,%
\text{mod}\left[ k\right] }^{1\sigma }\right \rangle +\exp \left[ -\frac{%
i\pi k}{N^{\prime }}\right] \left \langle \psi _{n,\text{mod}\left[ k\right]
}^{1\sigma \ast }\psi _{n,\text{mod}\left[ k\right] }^{2\sigma }\right
\rangle \\
+\exp \left[ \frac{i\pi k}{N^{\prime }}\right] \left \langle \psi _{n,\text{%
mod}\left[ k\right] }^{2\sigma \ast }\psi _{n,\text{mod}\left[ k\right]
}^{1\sigma }\right \rangle +\left \langle \psi _{n,\text{mod}\left[ k\right]
}^{2\sigma \ast }\psi _{n,\text{mod}\left[ k\right] }^{2\sigma }\right
\rangle%
\end{array}%
\right \}  \notag \\
&=&\frac{1}{4}\left \{ g_{n,\text{mod}\left[ k\right] }^{11\sigma }+g_{n,%
\text{mod}\left[ k\right] }^{22\sigma }+\exp \left[ -\frac{2i\pi k}{N_{s}}%
\right] g_{n,\text{mod}\left[ k\right] }^{12\sigma }+\exp \left[ \frac{2i\pi
k}{N_{s}}\right] g_{n,\text{mod}\left[ k\right] }^{21\sigma }\right \} \text{%
,}  \notag
\end{eqnarray}%
where $\text{mod}\left[ k\right] =\text{mod}\left[ k,N^{\prime }\right] =%
\text{mod}\left[ k,N_{s}/2\right] $.

Substituting the solution of the gap equation, one finally obtains, for half
filling:%
\begin{equation}
g_{nk}^{sym}=\frac{i\omega _{n}+2\cos \left[ \frac{2\pi }{N_{s}}k\right] }{%
\omega _{n}^{2}+e_{\text{mod}\left[ k\right] }^{2}}\text{.}  \label{AF_sym}
\end{equation}%
An example of results is compared with exact diagonalization for $%
N_{s}=4,T=0.2$ for $n=0$ and $k=0,N_{s}/4$ (corresponding to physical
momentum $\pi /2$) in Figs. 2 and 3a respectively. The symmetrized broken
phase solution (the dark green curve) for $U>U_{c}\left( T\right) \ $%
provides a quite accurate approximant. It approaches the exact result at
large coupling although still incorrectly indicates the second order
transition (see the cusps in inserts of both Figs. 2 and 3). The most
problematic values of both the frequency, $\omega =\pi T$ and the
quasimomentum $k=0$ and $N_{s}/4$ (Fermi surface) are chosen. The other
physical quantities are discussed in Section VI.

A question arises. Since qualitative features are captured quite well by the
symmetrized CGA except near the spurious transition, can one improve upon
this using the CGA as a starting point of a perturbation? This is attempted
next.

\section{Perturbative improvement of the gaussian theory}

\subsection{General construction of the series}

The covariant gaussian approximation can serve as a starting point for a
perturbation theory around the Hartree - Fock solution. In bosonic models
the method was proposed in the context of strong thermal fluctuations in the
mixed state of superconductor under magnetic field\cite{Thouless}. One
considers the quadratic form,%
\begin{equation}
\mathcal{A}_{g}=-\psi _{a}^{A\ast }\left[ G^{-1}\right] _{b-a}^{BA}\psi
_{b}^{B}\text{,}  \label{quadratic}
\end{equation}%
as a ``large" part of the action, while the difference between the models
action Eq.(\ref{action}) and it is ``small". The small part is multiplied by
a parameter $\alpha $ and the physical quantity is expanded in $\alpha $ to
a certain order. After the calculation is completed one sets $\alpha =1$.

The gaussian action for an interacting electron system is (as before the
space index combined with time):

\begin{eqnarray}
\mathcal{A} &=&\mathcal{A}_{g}+\alpha \Delta \mathcal{A}\text{;}
\label{delA} \\
\Delta \mathcal{A}\text{ } &=&\frac{1}{2}\psi _{a}^{A\ast }\psi
_{a}^{A}V_{a-b}^{AB}\psi _{b}^{B\ast }\psi _{b}^{B}+\psi _{a}^{A\ast }\left(
T_{a-b}^{AB}+\left[ G^{-1}\right] _{b-a}^{BA}\right) \psi _{b}^{B}\text{.}
\notag
\end{eqnarray}%
Integrands in the path integral are expanded as%
\begin{equation}
e^{-\left( \mathcal{A}_{g}+\alpha \Delta \mathcal{A}\right) }=e^{-\mathcal{A}%
_{g}}\left( 1-\alpha \Delta \mathcal{A}+\frac{1}{2}\alpha ^{2}\left( \Delta
\mathcal{A}\right) ^{2}+..\right) \text{.}  \label{expansion}
\end{equation}%
The correlator therefore is expanded to $\alpha $ as%
\begin{eqnarray}
\left \langle \psi _{a}^{A\ast }\psi _{b}^{B}\right \rangle &=&\frac{\int
\psi _{a}^{A\ast }\psi _{b}^{B}e^{-\mathcal{A}_{g}}\left( 1-\alpha \Delta
\mathcal{A}+\frac{1}{2}\alpha ^{2}\left( \Delta \mathcal{A}\right)
^{2}\right) }{\int e^{-\mathcal{A}_{g}}\left( 1-\alpha \Delta \mathcal{A}+%
\frac{1}{2}\alpha ^{2}\left( \Delta \mathcal{A}\right) ^{2}\right) }
\label{GFexpansion} \\
&\approx &G_{ab}^{AB}+\alpha \left \langle \psi _{a}^{A\ast }\psi _{b}^{B}%
\text{ }\Delta \mathcal{A}\right \rangle _{con}+\frac{\alpha ^{2}}{2}\left
\langle \psi _{a}^{A\ast }\psi _{b}^{B}\left( \Delta \mathcal{A}\right)
^{2}\right \rangle _{con}  \notag \\
&=&G_{ab}^{AB}+\alpha ^{2}\Delta G_{ab}^{AB}\text{.}  \notag
\end{eqnarray}%
The $\left \langle ...\right \rangle _{con}$ average is understood in a
diagrammatic representation of the gaussian integrals as in perturbation
theory\cite{NO} (division by $Z$ eliminates disconnected diagrams).
Vanishing of the $\alpha $ term is tantamount to solution of the gap
equation, Eq.(\ref{gapeq}), as shown in ref.\onlinecite{Thouless} (no difference
here between bosonic and fermionic models). $G_{ab}^{AB}$ is the green
function of Gaussian (Hartree-Fock) approximation.

The correction to the Gaussian correlator $G_{ab}^{AB}$ is (setting $\alpha
=1$ and simplifying by repeated use of the gap equation)%
\begin{equation}
\Delta G_{ab}^{AB}=G_{ak}^{AK}V_{kl}^{KL}G_{nl}^{NL}\left(
G_{kn}^{KN}G_{lm}^{LM}-G_{km}^{KM}G_{ln}^{LN}\right) V_{mn}^{MN}G_{mb}^{MB}%
\text{.}  \label{delG}
\end{equation}%
It is known that within gaussian approximation the effective action is
calculated much more precisely compared to correlators\cite{Kleinert}. The
cumulant ( the inverse of the green function that is the second functional
derivative of the effective action with respect to field) within the first
order is given by a simpler formula:%
\begin{eqnarray}
\left[ \left( G^{pg}\right) ^{-1}\right] _{ab}^{AB} &=&\left[ G^{-1}\right]
_{ab}^{AB}+\Sigma _{ab}^{AB};  \label{delGinv} \\
\Sigma _{ab}^{AB} &=&V_{al}^{AL}G_{nl}^{NL}\left(
G_{ab}^{AB}G_{ln}^{LN}-G_{an}^{AN}G_{lb}^{LB}\right) V_{bn}^{BN}\text{.}
\notag
\end{eqnarray}%
Here $\Sigma _{ab}^{AB}$ is self energy correction to the Gaussian
correlator, and $G^{pg}$ is the green function of post (perturbative
correction) gauss approximation (PCGA). Diagrammatically it can be
represented as summation of all the \textquotedblleft setting sun" diagrams
with lines representing the gaussian correlators, see Fig. 4.

\begin{figure}[tbp]
\begin{center}
\includegraphics[width=10cm]{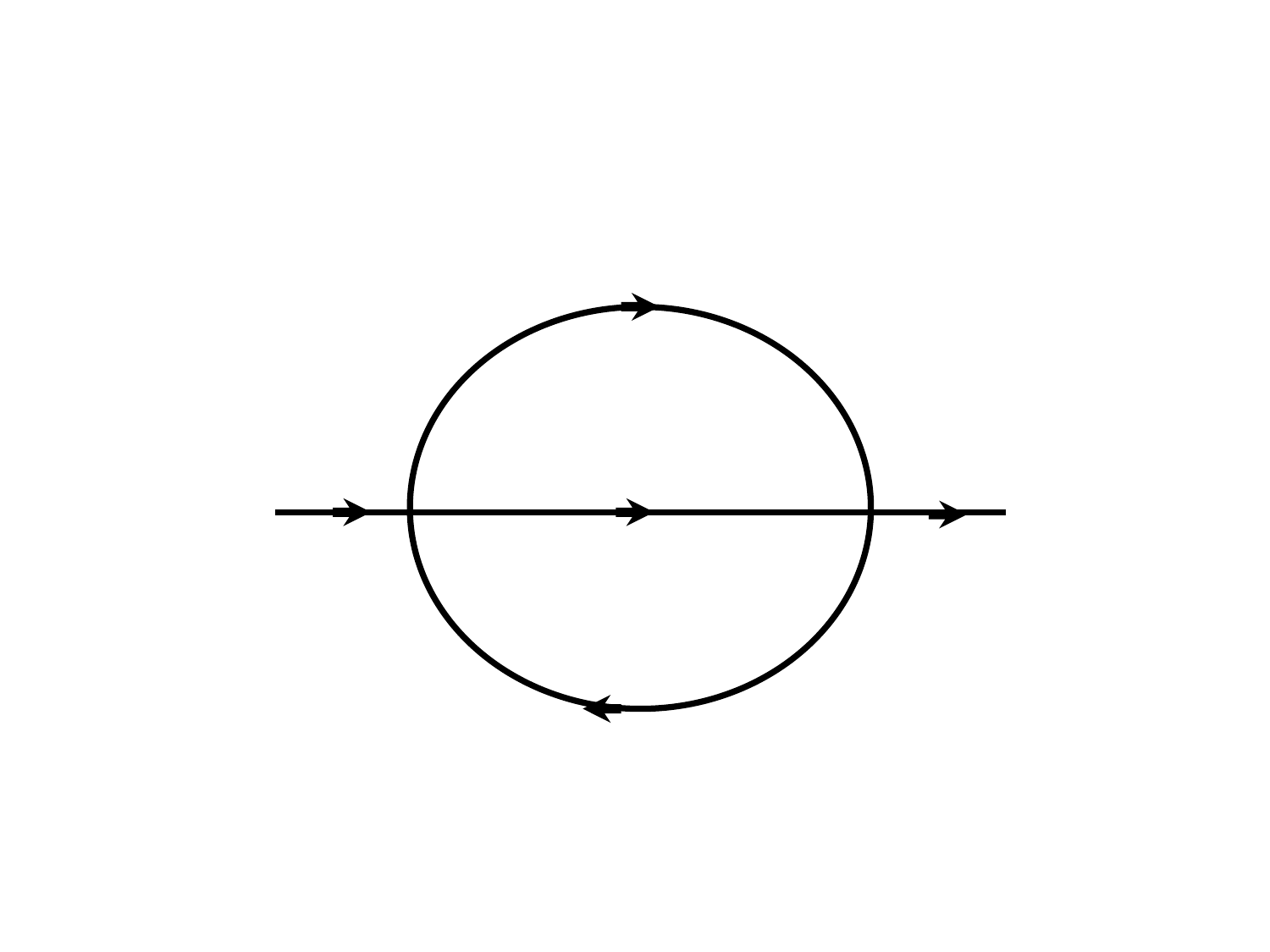}
\end{center}
\par
\vspace{-0.5cm}
\caption{Setting sun diagrams that contribute to PCGA. The directed lines
are gaussian correlators, while the vertices are ``perturbative".}
\end{figure}

As an example we calculate the first correction (``perturbed" or ``setting
sun" approximation) to the toy model of Section II. For the QD model,
substituting Eq.(\ref{intHubbard}) into Eq.(\ref{delGinv}), one obtains
(using the property of both the paramagnetic and the ferromagnetic solutions
that $G^{AB}$ is diagonal in spin due to the residual $U\left( 1\right) $
symmetry),%
\begin{equation}
\Sigma _{ab}^{AA}=\tau ^{2}U^{2}G_{ba}^{\overline{A}\overline{A}%
}G_{ab}^{AA}G_{ab}^{\overline{A}\overline{A}}\text{.}  \label{invQD}
\end{equation}%
Transforming to frequencies, one obtains,%
\begin{equation}
\Sigma _{n}^{AA}=T^{2}U^{2}\sum \nolimits_{k,l}g_{-n+k+l}^{\overline{A}%
\overline{A}}g_{k}^{AA}g_{l}^{\overline{A}\overline{A}}\text{,}
\label{invkQD}
\end{equation}%
so that in paramagnet, Eq.(\ref{QDexact}), for infinite $N_{t}$, $\left[
\gamma _{CGA}^{-1}\right] ^{AB}=\delta ^{AB}g^{CGA}$ with%
\begin{equation}
g_{n}^{CGA}=i\left \{ \omega _{n}+U^{2}T^{2}\sum \nolimits_{k,l=-\infty
}^{\infty }\frac{1}{\omega _{k+l-n}\omega _{k}\omega _{l}}\right \} ^{-1}%
\text{.}  \label{pgQD}
\end{equation}

The sum can be performed resulting in the exact expression given in Eq.(\ref%
{QDexact}). The calculation of the setting sun correction in the magnetic
phase is more complicated, however the result is simple (after
symmetrization):%
\begin{equation}
g_{n}^{pert}=i\omega _{n}\frac{U^{2}/4+\omega _{n}^{2}}{\left( 2UM\omega
_{n}\right) ^{2}+\left( 2U^{2}M^{2}-U^{2}/4-\omega _{n}^{2}\right) ^{2}}%
\text{.}  \label{PCGA_QD}
\end{equation}%
Here magnetic moment $M$ is determined by the gap equation Eq.(\ref{ferroeq}%
). The correlator of Eq.(\ref{PCGA_QD}) is plotted as the purple line in
Fig. 1. The most difficult case of $\omega =\pi T$ is given in
Fig.1. One observes that it significantly improves the symmetrized CGA near
the spurious transition at $U_{c}$ (see inset), but is not effective at
higher couplings. If $U<U_{c}$, the perturbative correction turns out to be
exact. The asymptotics for large coupling is correct and corrections are
exponential. The improvement is dramatic for larger frequencies, as can be
seen from Fig.1b.

\subsection{Perturbative correction to gaussian approximation in the Hubbard
model}

Applying the general formula for the setting sun corrected self energy, Eq.(%
\ref{delGinv}) in the anti - ferromagnetic phase of the Hubbard model, one
obtains:

\begin{eqnarray}
\Sigma _{\alpha }^{\upharpoonleft IJ} &=&\frac{T^{2}U^{2}}{N^{\prime }}\sum
\nolimits_{\chi _{1},\chi _{2}}G_{\chi _{1}}^{\upharpoonleft IJ}G_{\chi
_{2}}^{\downarrow IJ}G_{\chi _{1}+\chi _{2}-\alpha }^{\downarrow JI}\text{,}
\label{AF_Sigma} \\
\Sigma _{\alpha }^{\downarrow IJ} &=&\frac{T^{2}U^{2}}{N^{\prime }}\sum
\nolimits_{\chi _{1},\chi _{2}}G_{\chi _{1}}^{\downarrow IJ}G_{\chi
_{2}}^{\upharpoonleft IJ}G_{\chi _{1}+\chi _{2}-\alpha }^{\upharpoonleft JI}
\notag
\end{eqnarray}%
where $I,J$ are sublattice indices, $\alpha ,\chi $ indices are the combined
indices of frequency and wavevector. Substituting the HF anti -
ferromagnetic solution of Eq.(\ref{algebraic_gapeq}) in the matrix form the
correlator is:

\begin{equation}
G_{nk}^{\sigma }=\frac{1}{\omega _{n}^{2}+4\cos ^{2}\left[ \frac{2\pi k}{%
N_{s}}\right] +M^{2}U^{2}}%
\begin{pmatrix}
i\omega _{n}+\left( -\right) ^{\sigma }MU & 1+\exp \left[ 4i\pi k/N_{s}%
\right] \\
1+\exp \left[ -4i\pi k/N_{s}\right] & i\omega _{n}-\left( -\right) ^{\sigma
}MU%
\end{pmatrix}%
\text{.}  \label{sigma1}
\end{equation}%
where $\sigma $ is the spin index, and for spin up $\sigma =1$, spin down $%
\sigma =2$. Using Eq.(\ref{delGinv}), the PCGA correlators are obtained and
the symmetrization of the PCGA correlators follows. The symmetrized PCGA
correlators are plotted in the different figures of the present paper using
purple lines or points. The generalization to higher dimensions, different
dispersion relations/lattices, beyond half filling etc is straightforward.

These results are systematically compared with exact and Monte Carlo
simulations in the 1D Hubbard model in the next section and with 2D Hubbard
model in Section V.

\section{Comparison with exact diagonalization and the Monte Carlo
simulation of the Hubbard Model}

Exact solutions of strongly interacting electron systems are scarce. This
especially true for Green's function at finite temperature. We use exact
diagonalization\cite{ED} in 1D for small lattice at any filling (standard
and thus not described here) and then utilize the determinant quantum Monte
Carlo\cite{dqmc} (DQMC, briefly described in Appendix) for half filling
only. Although the methodology has been extended recently to approach
electronic systems beyond the half filling, for the benchmarks purpose we
stay with well established half filling domain for which the sign problem
was shown to be nonexistent.

\subsection{Coupling and quasi - momentum dependence of the Green's function
of half filled Hubbard chain}

In this subsection our analytic results are compared with exact
diagonalization of the 1D half filling in the most troublesome case of half
filling (appears as red lines in figures). Results beyond half filling are
in far better agreement with exact even for deviation as small as $\delta
\mu =\mu -U/2=0.1$. At half filling, the imaginary part of the Green's
function at quasimomentum in the $\Gamma $ point, $k=0$, and on the Fermi
surface $k=1$ (corresponding to the physical wave vector $\pi /2$) is shown
on Fig. 2 and Fig.3a respectively for $N_{s}=4$. The results are for fixed
temperature (in units of hopping energy $t=1)$, $N_{s}=4$, and at lowest
Matsubara frequency $\omega _{n=0}=\pi T$ (by far the most difficult case,
as example of the simpler model demonstrates, see Fig.1). The range of
couplings $U_{c}<U<12$ is shown with inset magnifying the region around the
spurious critical value $U_{c}=1.3335$ marked by the dashed line for $T=0.2$
in Fig. 2 and Fig. 3a for $T=0.2$, $N_{s}=4$. All the calculations here are
for infinite $N_{t}$. In Figs. 3a, 3b, 3c imaginary part of the Green's
function at quasimomentum $k=1$ (corresponding to the physical wave vector $%
\pi /2$) is shown respectively for $\delta \mu =0$, $\delta \mu =0.25$ and $%
\delta \mu =1$.

Below the spurious phase transition HF (green straight segment) deviates
significantly from the exact diagonalization result (red curve), especially
near $U_{c}$. However well above $U_{c}$ the symmetrized CGA result (green
curve) compares well with the exact correlator. On the Fermi surface, $k=\pi
/2$, the perturbative improvement over the symmetrized CGA (the purple curve
in Figs.2, 3a) is significant not just near the spurious transition at $%
U_{c} $, but all the way to the large $U$ limit. The leading large $U$
asymptotic, $g=\frac{4\pi T}{U^{2}}$($\frac{2.51}{U^{2}}$for $T=0.2$) is
captured correctly by both CGA and the perturbatively improved CGA for both
quasimomentum $k=0,\pi /2$. However the coefficient $c$ of the subleading, $%
c/U^{4}$, correction (powers are even due to the particle - hole symmetry)
is different. The exact one for $k=0$ is $c=40.6$, while approximate are $%
c=-24.1$ and $c=-8.5$ for CGA and the perturbatively corrected CGA (PCGA)
respectively. For $k=\pi /2$ the situation is similar: exact $c=86.0$, while
CGA and PCGA give $c=29.8$ and $c=-27.8$ correspondingly. The conclusion is
that for very strong antiferromagnetic state the dominant correlation is
antiferromagnetic and the long range symmetrization is less important. The
perturbation thus is not helpful in this respect. Its main advantage is at
intermediate couplings. The most important positive observation is that
symmetrized mean field works better beyond half filling.

However, in the case of QD, the large $U$ limit expansion (polynomial
expansion $U^{-2k},k=1,2..$) of the correlator from Eq.(\ref{PCGA_QD}) is
the same as the exact one, and the difference between them is exponential
small factor ($e^{-0.5U}$).

For large $N_{s}$ the exact diagonalization is impossible and thus DQMC was
used as a benchmark. We present next comparison of the quasimomentum
distribution for large enough chain, so that the continuum limit is reached.
\begin{figure}[tbp]
\begin{center}
\includegraphics[width=10cm]{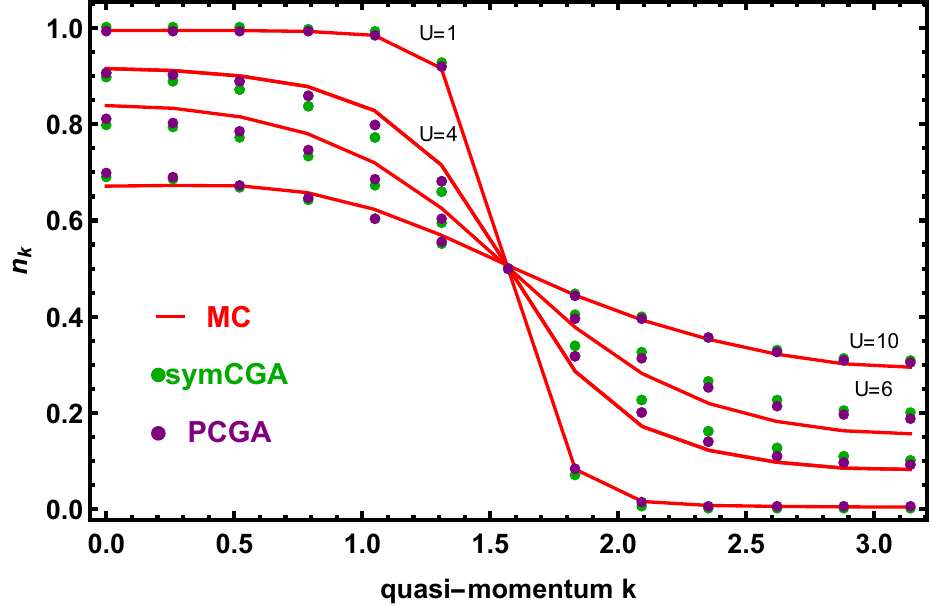}
\end{center}
\par
\vspace{-0.5cm}
\caption{The quasi - momentum distribution function $n_{k}=2T\sum%
\limits_{m}g\left( m,k\right) $ in half filled 1D Hubbard model ($N_{s}=24$%
). factor $2$ is due to spin summation. horizontal axis as $k$ is
quas-momentum, and quantized as in unit $2\protect \pi /N_{s}$, and the plot
range of $k$ is from $0$ to $\protect \pi $. The results for CGA (green
dots), and PCGA (purple dots) are plotted along with MC results (red lines).
}
\end{figure}

\subsection{Distribution of momenta in 1D Hubbard model}

In Fig. 5 the coupling dependence of the distribution function $n_{k}$ of
the $N_{s}=24$ Hubbard chain are compared with determinant quantum Monte
Carlo simulation (red line), see Appendix for details. Temperature is again
fixed at $T=0.2$, while couplings are $U=1,4,6,10$. The spurious transition
occurs at $U_{c}=2.017$ very close to the value mean field transition point $%
U_{c}=2.0186$ in the thermodynamic limit $N_{s}=\infty $, so that it
essentially represents the continuum limit. We use the infinite $N_{t}$
limit for the symmetrized HF (green points) and PCGA (purple points).

One observes that at the weak coupling ($U=1$) the agreement is excellent
and the perturbation improves significantly the gaussian result. The weak
coupling limit comparison means that the MC simulation time slice
corresponding to $N_{t}=40$ is precise enough. For an intermediate coupling
above $U_{c}$ ($U=4,6$) there are deviations of up to $10\%$ at certain
momenta, that are only modestly corrected perturbatively. Finally at strong
coupling ($U=10$) the agreement is good, but the perturbative correction
does not help much.

\subsection{Charge and spin correlators in 1D Hubbard model}

In this subsection more complicated correlators of the Fermionic fields are
compared with exact results on small lattice and Monte Carlo simulations of
the half filled model on larger ones.

In Fig. 6 the coupling dependence of the charge density correlator $\chi
_{n,k}^{\rho }=\left \langle n_{n,k}n_{-n,-k}\right \rangle $ of the $N_{s}=4$
Hubbard chain at half filling is compared with the exact diagonalization
(red line). The subindices $n,k$ of $n_{n,k}$ corresponds to Matsubara
frequency $\omega _{n}=\pi T\left( 2n+1\right) $, $k$ is the quasi-momentum,
$n_{n,k}$ is the Fourier transformations of the density $n_{\tau ,x}$.
Temperature is fixed at $T=1$, frequency at $n=0$, momentum at $k=0$, while
the coupling range is $U=0-12$. The spurious transition occurs at $%
U_{c}=4.541$. The symmetrized density correlator (the customary Lindhard
diagram with propagators given by the HF approximation, green lines)
deviates from exact result near the spurious transition, although it has a
correct asymptotics at both weak and strong couplings.

The one vertex corrected symmetrized density correlator (purple points) does
better. It is within 1\% in the ``unbroken" phase (including the spurious
transition point), and improves the intermediate region. Inset demonstrates
the ration of an approximate and the exact correlator at large coupling.

\begin{figure}[tbp]
\begin{center}
\includegraphics[width=10cm]{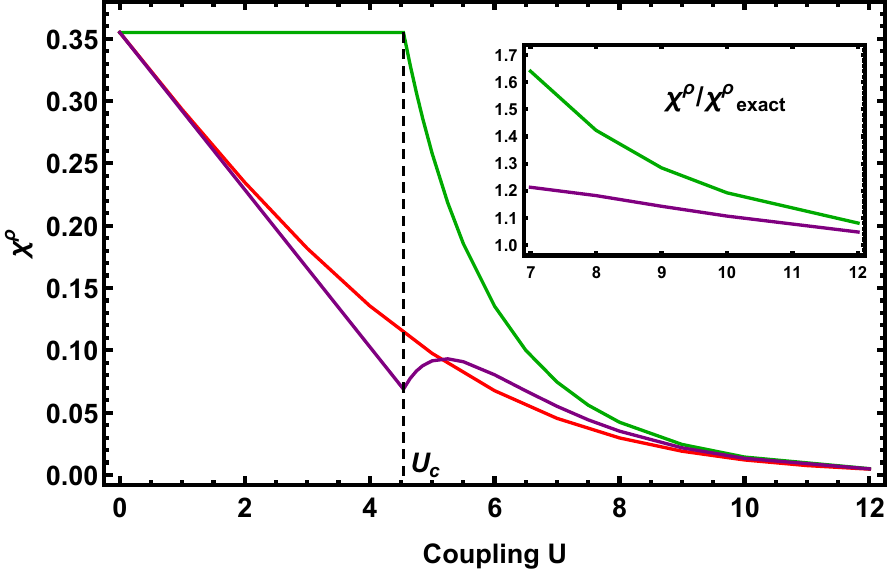}
\end{center}
\par
\vspace{-0.5cm}
\caption{The charge density correlator $\protect \chi ^{\protect \rho }=$ $%
\protect \chi _{n,k}^{\protect \rho }$ dependence of $U$ at $T=1.$, for frequency
 $n=0$, momentum at $k=0$. The green curve is the  Lindhard
diagram result, and the purple curve contains in addition the next order
correction to the Lindhard diagram. The green curve and the purple curve in the
inset are the ratios between the Lindhard result and the result including
the next order correction to the exact value respectively.}
\end{figure}

\begin{figure}[tbp]
\begin{center}
\includegraphics[width=10cm]{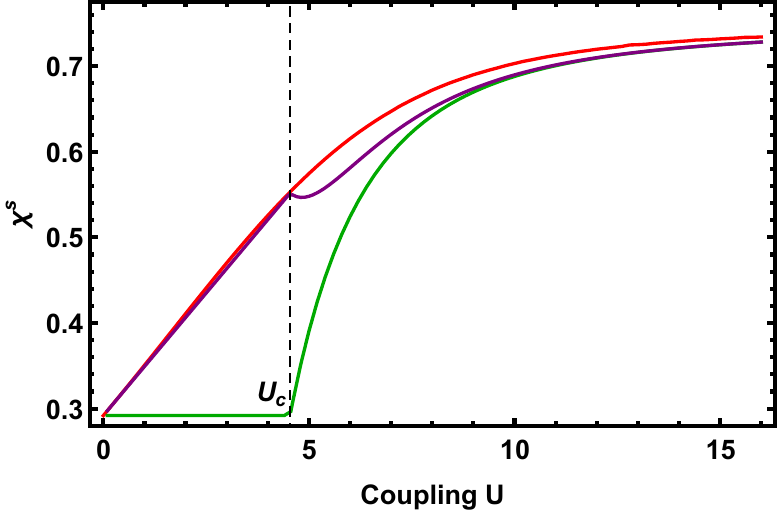}
\end{center}
\par
\vspace{-0.5cm}
\caption{The spin correlator $\protect \chi _{n}^{s}=\left \langle \protect%
\overrightarrow{S}_{n,x}\cdot \protect \overrightarrow{S}_{-n,x}\right
\rangle $ dependence of $U$ at $T=1$ for frequency  $n=0$. The green curve is
the Lindhard diagram result, and the purple curve contains  in addition
 the next order correction to the Lindhard diagram.}
\end{figure}

Another interesting correlator is the spin correlator $\chi _{n,k}^{s}=\left
\langle \overrightarrow{S}_{n,k}\cdot \overrightarrow{S}_{-n,-k}\right
\rangle $. The subindices $n,k$ of $\overrightarrow{S}_{n,k}$ corresponds to
Matsubara frequency $\omega _{n}=\pi T\left( 2n+1\right) $, $k $ is the
quasi-momentum, and $\overrightarrow{S}_{n,k}$ is the Fourier
transformations of the spin $\overrightarrow{S}_{\tau ,x}$. \ Parameters $%
N_{s}$, $T$ are the same as for the density correlator, frequency still at $%
n=0$, but instead of quasimomentum we take the coincident point correlator $%
\chi _{n}^{s}=\left \langle \overrightarrow{S}_{n,x}\cdot \overrightarrow{S}%
_{-n,x}\right \rangle $ for $n=0$. The results are presented in Fig. 7 as
function of the coupling in the range $U=0-16$. The approximation quality is
approximately (a little worse than density correlator) the same as in the
previous case of the density correlator. For results of large $N_{s}$, we
will present the results in the future works. \

\begin{figure}[tbp]
\begin{center}
\includegraphics[width=10cm]{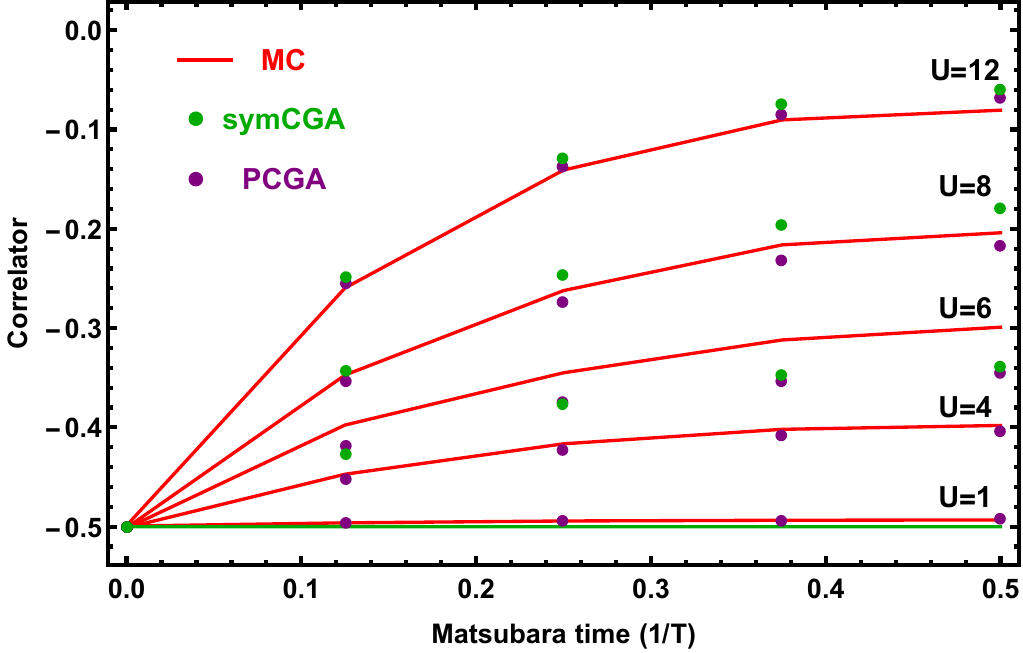}
\end{center}
\par
\vspace{-0.5cm}
\caption{Imaginary part of the Green's function $G\left( \protect \tau ,%
\mathbf{k}\right) $ of half filled 2D Hubbard model at $k=\left( \protect \pi %
,0\right) $. The results for CGA (green dots), and PCGA (purple dots) are
plotted along with MC results (red lines). }
\end{figure}

\subsection{Comparison of MC simulation with CGA in 2D Hubbard model}

Calculations for half filled Hubbard model in 2D are completely analogous to
those in 1D. In Fig. 8 the coupling dependence of the Matsubara Green's
function at the point $k=\left( \pi ,0\right) $ on the Fermi surface is
plotted as function of Matsubara time. As was demonstrated in the previous
subsection, momenta on the Fermi surface are most difficult to describe. The
temperature is fixed at $T=1$, and only half of the period $0<t<1/\left(
2T\right) $ is shown since the other half is dictated by the symmetry. The
number of space points was $144$ with $N_{s}=12$ in the DQMC simulation (red
line), while the time slice corresponds to $N_{t}=8$, see more detailed
description of methodology in Appendix. The couplings, $U=1,4,6,8,12$, were
taken below and above the spurious mean field transition at $U_{c}=4.90$. We
use the infinite $N_{t}$ limit for the symmetrized HF (green points) and
PCGA (purple points).

One observes that at the weak coupling ($U=1,4$) below $U_{c}$ the agreement
is excellent only if the HF (the green vertical line) is perturbatively
corrected as in 1D. The weak coupling limit comparison means that the MC
simulation time slice corresponding to $N_{t}=8$ is precise enough. For an
intermediate coupling just above $U_{c}$ ($U=6$) there are significant
deviations of up to $15\%$ , that are not corrected perturbatively. Finally
at a stronger couplings ($U=8,12$) the agreement is good, but improvement
(perturbative correction) does not help much.

\section{Discussion and conclusions}

To summarize, a mean field (Hartree-Fock) type approach, covariant gaussian
approximation is adapted to include strongly interacting low dimensional
electronic systems in which symmetry is \textquotedblleft restored" due to
long range correlations. Instead of using a complicated (typically
renormalization group type) scale separation method, simple symmetrization
of correlators is employed to a covariant (preserving Ward - Takahashi
identities) variant of the mean field (gaussian) approximation. The short
range correlations captured by the mean field are thus kept, while symmetry
gets restored. The solution can be systematically improved by addition of
corrections to cumulant that are based on expansion around the gaussian
approximation. There are different \ variational Hartree-Fock methods \cite%
{Tomita} which were applied to study the strong correlated model, for
example, the Hubbard model with success. However here we offered the
traditional (simple analytic) Hartree-Fock methods to calculate the
correlators.

To test the scheme, it was applied to the\ correlator of the 1D and the 2D
one band Hubbard models and compared to exact diagonalization of relatively
small systems (ED) and MC simulations at the half filling, where they are
known to be reliable. The comparison demonstrates the typical mean field
precision of order 10\% for all couplings. It is better for weak and strong
couplings (correct asymptotics) away from the Fermi level and higher
frequencies. It should be noted that the method generalizes well beyond half
filling Hubbard model. The 2D Hubbard model beyond the half filling that is
being intensely studied recently in connection to strange metals and high $%
T_{c}$ superconductivity (including by the determinantal quantum Monte Carlo%
\cite{dqmc} used here at half filling only). Apart from straightforward
generalizations to different symmetry groups describing for example Ising or
XY quantum magnets, possible applications include models describing phonon
induced interactions like the Holstein model. For disordered matter, one can
combine replica field theory method with the method used in the present
paper.

A natural question arises whether the symmetrization scheme can be applied
to finer approximations beyond the gaussian. Recently the covariant
approximation method was generalized to include higher cumulants beyond the
quadratic\cite{CCA}.

\textit{Acknowledgements}

Authors are very grateful to J. Wang, B. Shapiro, for numerous discussions
and I. Berenstein and G. Leshem for help in computations. B. R. was
supported by MOST of Taiwan \#107-2112-M-003-012. D. P. L. was supported by
National Natural Science Foundation of China (No. 11674007 and No.
91736208). B. R. and D.P.L are grateful to School of Physics of Peking
University and The Center for Theoretical Sciences of Taiwan for hospitality
respectively.

\appendix

\section{Determinant Quantum Monte Carlo}

The determinant quantum Monte Carlo (DQMC) method is an exact numerical tool
to treat the correlated system. To apply DQMC simulations in fermion system,
a major obstacle is the notorious sign problem, which prevents DQMC
simulations from achieving a good numerical precision at low temperature and
high interaction strength. However, in the half-filled Hubbard model on a
square lattice, the sign problem disappears due to the particle hole
symmetry, and this provide a wonderful opportunity to use the data of DQMC
as the benchmark for our method.

The DQMC method that we use is based on Blankenbecler--Scalapino--Sugar
(BSS) algorithm\cite{dqmc}. In this Appendix, we present a brief
introduction following previous work on the Hubbard model\cite{Hirsch}. The
Hamiltonian Eq.(\ref{HubbardH}) can is separated into $H=H_{0}+H_{I}$ where $%
H_{0}$ is the hopping part and $H_{I}$ is includes the rest of terms in Eq.(%
\ref{HubbardH}). In order to calculate the grand partition function $Z=Tr\
e^{-H/T}$, one need to use the Suzuki-Trotter decomposition scheme\cite%
{Suzuki}, to cast the quartic term into a bilinear form, and introduce a
small parameter $\tau =\left( TN_{t}\right) ^{-1}$,%
\begin{equation}
e^{-(H_{0}+H_{I})/T}=\left( e^{\tau H_{0}+\tau H_{I}}\right) ^{N_{t}}=\left(
e^{\tau H_{0}}e^{\tau H_{I}}\right) ^{N_{t}}+O\left( \tau ^{2}U\right) .
\label{im-time}
\end{equation}

Having separated the exponentials, we can decouple the quartic terms in $%
H_{I}$ by the Hubbard--Stratonovich (HS) transformation,
\begin{equation}
e^{-U\tau n_{\uparrow }n_{\downarrow }}=\frac{1}{2}e^{-\frac{U\tau }{2}%
n}\sum_{s=\pm 1}e^{-s\xi \left( n_{\uparrow }-n_{\downarrow }\right) }=\frac{%
1}{2}\sum_{s=\pm 1}\prod_{\sigma =\uparrow ,\downarrow }e^{-\left( sgn\left[
\sigma \right] s\xi +\frac{U\tau }{2}\right) n_{\sigma }}\text{,}  \label{HS}
\end{equation}%
where $n\equiv n_{\uparrow }+n_{\downarrow }$. and the parameter $\xi =$
arccosh$\left[ e^{|U|\tau /2}\right] $. One can notice that the quartic
terms are decoupled at the cost of introducing an auxiliary field at every
site and time slice. Upon replacing the on-site interaction on every site of
the space-time lattice by Eq. (\ref{HS}), we obtain the sought of form in
which only bilinear terms appear in the exponential,
\begin{equation}
{\mathcal{Z}}=\left( \frac{1}{2}\right) ^{N_{s}^{D}N_{t}}\underset{\{s\}}{%
\text{{\Large \textrm{Tr}}}}\prod_{t=1}^{N_{t}}\prod_{\sigma =\uparrow
,\downarrow }\exp \left[ -\tau H_{0}\right] \exp \left[ -\tau \sum_{\mathbf{i%
}}c_{\mathbf{i}}^{\sigma \dagger }V_{\mathbf{i}}^{\sigma }(t)c_{\mathbf{i}%
}^{\sigma }\right]  \label{Z_MC}
\end{equation}%
where the traces are over auxiliary Ising fields and over fermion
occupancies on every site. The time-slice index $t$ is manifested in the HS
field $s_{i}(t)$ by
\begin{equation}
V_{i}^{\sigma }(t)=sgn\left[ \sigma \right] \frac{\xi }{\tau }s_{i}(t)+\mu -%
\frac{U}{2}\text{,}  \label{Vi}
\end{equation}%
which are the elements of the $N_{s}\times N_{s}$ diagonal matrix $V^{\sigma
}(t)$. With bilinear forms in the exponential, the fermions can be traced
out explicitly,
\begin{equation}
{\mathcal{Z}}=\left( \frac{1}{2}\right) ^{N_{s}^{D}N_{t}}\underset{\{s\}}{%
\text{{\Large \textrm{Tr}}}}\, \prod_{\sigma }\det \left[ {\mathbf{1}}+%
\mathsf{B}^{\sigma }\left( N_{t}\right) \mathsf{B}^{\sigma }\left(
N_{t}-1\right) \ldots \mathsf{B}^{\sigma }\left( 1\right) \right] ,
\end{equation}%
with $\mathsf{B}^{\sigma }\left( t\right) \equiv e^{-\tau K}e^{-\tau
V^{\sigma }(t)}$, in which the auxiliary Ising spins are implicitly
included. The hopping terms in the exponential are represented by an $%
N_{s}\times N_{s}$ matrix $K$, with elements
\begin{equation}
K_{ij}=%
\begin{cases}
-1 & \text{if $i$ and $j$ are nearest neighbours}, \\
0 & \text{otherwise}.%
\end{cases}
\label{Kij}
\end{equation}

The equal-`time' correlation function of the creation and the annihilation
operators is:
\begin{equation}
\langle c_{\mathbf{i}}^{\sigma }c_{\mathbf{j}}^{\sigma \dagger }\rangle =%
\frac{1}{\mathcal{Z}}\, \underset{\{s\}}{\text{{\Large \textrm{Tr}}}}T{\text{%
{\Large $r$}}}\left[ c_{\mathbf{i}}^{\sigma }c_{\mathbf{j}}^{\sigma \dagger
}\prod_{t,\sigma }e^{-\tau K}e^{-\tau V^{\sigma }(t)}\right] \text{.}
\label{AB}
\end{equation}%
Considering the fact that the fermions only interact with the auxiliary
fields, it can be proved that Wick's theorem \cite{FW} holds \textit{for a
fixed HS configuration} \cite{Hirsch,vdl92,Loh92}. Hence, the interesting
physical expectations can be calculated in terms of single-particle Green's
functions. In the `Heisenberg picture', the time-dependent $c$ operator is
defined as,
\begin{equation}
c(t)\equiv e^{t\tau H}\,c\,e^{-t\tau H}\text{,}  \label{heispic}
\end{equation}%
with the initial time set to be $t=\tau $ and $c^{\dagger }(t)\neq \lbrack
c(t)]^{\dagger }$. Further, the unequal-time Green's function, for $%
t_{1}>t_{2}$, is given by \cite{Hirsch}
\begin{eqnarray}
G_{\mathbf{i}\mathbf{j}}^{\sigma }(t_{1};t_{2}) &\equiv &\left \langle c_{%
\mathbf{i}}^{\sigma }(t_{1})c_{\mathbf{j}}^{\sigma \dagger }(t_{2})\right
\rangle _{\{s\}}  \notag \\
&=&\left[ \mathsf{B}^{\sigma }\left( t_{1}\right) \mathsf{B}^{\sigma }\left(
t_{1}-1\right) \ldots \mathsf{B}^{\sigma }\left( t_{2}+1\right) \, \mathsf{g}%
^{\sigma }(t_{2}+1)\right] _{\mathbf{i}\mathbf{j}}\text{,}  \label{gtau}
\end{eqnarray}%
in which the Green's function matrix at the $t$-th time slice is defined as $%
\mathsf{g}^{\sigma }(t)\equiv \left[ \mathsf{1}+\mathsf{A}^{\sigma }(t)%
\right] ^{-1}$ with $\mathsf{A}^{\sigma }(t)\equiv \mathsf{B}^{\sigma
}\left( t-1\right) \mathsf{B}^{\sigma }\left( t-2\right) \ldots \mathsf{B}%
^{\sigma }\left( 1\right) \mathsf{B}^{\sigma }\left( N_{t}\right) \ldots
\mathsf{B}^{\sigma }\left( t\right) $.

In our simulations, $8000$ sweeps were used to equilibrate the system. An
additional $30000$ sweeps were then made, each of which generated a
measurement. These measurements were split into ten bins which provide the
basis of coarse-grain averages and errors estimates based on standard
deviations from the average. In the determinant QMC method, a breakup of the
discretized imaginary time evolution operator introduces a systematic error
proportional to $\tau ^{2}U$ (with $\tau =\left( TN_{t}\right) ^{-1}$ being
the imaginary time step). We have used $\tau =0.125$, which leads to
negligible systematic error (within a few percent). One of the authors had
succeeded in using this technology to explore interesting physical
properties in various electronic systems \cite{Maqmc}.


\begin{thebibliography}{99}
\bibitem{Mermin} N.D. Mermin and H. Wagner, Phys. Rev. Lett. \textbf{17},
1133 (1966).

\bibitem{Chaikin} P. M. Chaikin and T. C. Lubensky, \textquotedblleft
Principles of condensed matter physics", Campridge University Press, 1995.

\bibitem{Jevicki} A. Jevicki, Phys. Let. B \textbf{71}, 327 (1977).

\bibitem{David} F. David, Com. Math. Phys. \textbf{81}, 149 (1981). S.
Elitzur, Nucl. Phys. B\textbf{212} , 501 (1983).

\bibitem{Maki} K. Maki and H. Takayama, Prog. Theor. Phys. \textbf{46}, 1651
(1971).

\bibitem{Kao} B. Rosenstein, \ Phys. Rev B\textbf{60}, 4268, (1999); H.C.
Kao, B. Rosenstein and J.C. Lee, Phys. Rev. B\textbf{61}, 12352 (2000).

\bibitem{Lisolid} D. Li and B. Rosenstein, Phys. Rev.\textbf{B 65}, 024514
(2001).

\bibitem{Wang17} J. F. Wang, D. P. Li, H. C. Kao, and B. Rosenstein, Ann.
Phys. 380, 228 (2017).

\bibitem{CCA} B. Rosenstein and A. Kovner, Phys. Rev. \textbf{D40}, 523
(1989); B. Rosenstein and D. Li, Phys. Rev. \textbf{B98}, 155126 (2018).

\bibitem{NO} J. W. Negele and H. Orland, \textquotedblleft Quantum
Many-particle Systems", Perseus Books, 1998.

\bibitem{Haarmeasure} S. Steinberg, ``Group theory and physics", Cambridge
University Press (1994). S. Aubert and C. S. Lam, J. Math. Phys. \textbf{44}%
, 6112 (2003).

\bibitem{Rossi} P. Rossi, M. Campostrini, E. Vicari, Phys. Rep. \textbf{302}%
, 143 (1998); Z. Pucha\l a and J. A. Miszczak, Bulletin of the Polish
Academy of Sciences Technical Sciences \textbf{65}, 21 (2017).

\bibitem{Korepin} V. E. Korepin and F. H. L. Essler,\textquotedblleft
Exactly Solvable Models of Strongly Correlated Electrons", World Scientific,
1994.

\bibitem{Auerbach} A. Auerbach, ``Interacting electrons and quantum
magnetism". Springer Science \& Business Media, 2012.

\bibitem{Thouless} D.J. Thouless, Phys. Rev. Lett. \textbf{34}, 946 (1975);
G J Ruggeri and D J Thouless, J. Phys. \textbf{F} \textbf{6,} 2063 (1976);
S. Hikami, A. Fujita, and A. I. Larkin, Phys. Rev. \textbf{B 44}, 10400(R)
(1991); J. Hu, A. H. MacDonald, and B. D. McKay, Phys. Rev. B \textbf{49},
15263 (1994); B. Rosenstein and D. Li, Rev. Mod. Phys. \textbf{82}, 109
(2010).

\bibitem{Kleinert} H. Kleinert, \textquotedblright Path integrals in quantum
mechanics, statistics, and polymer physics\textquotedblright ,World
Scientific, Singapore (1995).

\bibitem{ED} A. Weisse and H. Fehske, \textquotedblleft Exact
Diagonalization Techniques", in \textquotedblleft Computational
Many-Particle Physics" edited by H. Fehske, R. Schneider, A. Weisse,
Springer, Berlin, 2008.

\bibitem{Tomita} N. Tomita, Phys. Rev. B \textbf{69}, 045110 (2004) and
references therein.

\bibitem{dqmc} R. Blankenbecler, D. J. Scalapino, and R. L. Sugar, Phys.
Rev. D \textbf{24}, 2278 (1981).

\bibitem{Hirsch} J. E. Hirsch, Phys. Rev. B \textbf{31}, 4403 (1985);
Raimundo R. dos Santos, Braz. J. Phys. \textbf{33}, 36 (2003); T. Ma, F. M.
Hu, Z. B. Huang, and H. Q. Lin, \textit{Horizons in World Physics.} \textbf{%
276}, Chapter 8, Nova Science Publishers, Hauppauge, New York, Inc. (2011).

\bibitem{Suzuki} \textit{Quantum Monte Carlo Methods,} Solid State Sciences,
Vol.\ 74, ed.\ M.\ Suzuki (Springer, Berlin), 1986.

\bibitem{FW} A.\ L.\ Fetter and J.\ D.\ Walecka, \textit{Quantum Theory of
Many-Particle Systems,} (McGraw-Hill, New York), 1971.

\bibitem{vdl92} W.\ von der Linden, Phys.\ Rep.\  \textbf{220}, 53 (1992).

\bibitem{Loh92} E.~Y. Loh and J.~E.~Gubernatis, in \textit{Electronic Phase
Transitions}, edited by W.~Hanke and Yu.~V.~Kopaev (Elsevier, Amsterdam,
1992).

\bibitem{Maqmc} T. Ma, H. Q. Lin, J. Hu, Phys. Rev. Lett. \textbf{110},
107002 (2013); S. Cheng, J. Yu, T. Ma, N. M. R. Peres, Phys. Rev. B \textbf{%
91}, 075410 (2015); G. Yang, S. Xu, W. Zhang, T. Ma, and C. Wu, Phys. Rev. B
\textbf{94}, 075106 (2016); T. Ma, L. Zhang, C.-C. Chang, H.-H. Hung, R. T.
Scalettar, Phys. Rev. Lett. \textbf{120}, 116601 (2018).
\end{thebibliography}
\end{document}